\documentclass[paper]{JHEP}
\usepackage{epsfig}
\usepackage{amsmath}
\usepackage{amssymb}
\def\beq{\begin{equation}}
\def\beqn{\begin{eqnarray}}
\def\eeq{\end{equation}}
\def\eeqn{\end{eqnarray}}
\def\abs#1{\left|#1\right|}

\def\bk{\bar k}

\def\bp{\bar p}

\def\bt{\bar t}

\def\bx{\bar x}

\def\dsb{d\bar\sigma}

\def \qb {\bar{q}}

\def \ap {{\cal A}_+}
\def \am {{\cal A}_-}
\def \real {\,\mathrm{Re}}
\def\bt{\bar t}
\newcommand\ab{\alpha\beta}
\newcommand\ampI{{\cal A}_{\ab}}
\newcommand\singlyI{{\cal S}_{\ab}}
\newcommand\doublyI{{\cal D}_{\ab}}
\newcommand\interfI{{\cal I}_{\ab}}
\newcommand\ssinglyI{\hat{\cal S}_{\ab}}
\newcommand\lumiI{{\cal L}_{\ab}}
\newcommand\wdoublyI{\widetilde{\cal D}_{\ab}}

\newcommand\interf{{\cal I}}
\newcommand\doubly{{\cal D}}
\newcommand\wdoubly{\widetilde{\cal D}}

\newcommand\sss{\scriptscriptstyle\rm}
\newcommand\pt{p_{\sss T}}
\newcommand\DR{{\sss\rm DR}}
\newcommand\DS{{\sss\rm DS}}
\newcommand\NI{{\sss\rm NI}}
\newcommand\LO{{\sss\rm LO}}
\newcommand\bq{\bar{q}}

\newcommand\fo{f_1}

\newcommand\xMCB{\Big|_{\sss {\rm MC}}}
\newcommand\Hone{(H_1)}
\newcommand\Htwo{(H_2)}

\newcommand\as{\alpha_{\sss S}}
\newcommand\gs{g_{\sss S}}
\newcommand\stepf{\Theta}
\newcommand\mydot{\!\cdot\!}

\newcommand\bb{\bar{b}}

\newcommand\mat{{\cal M}}

\newcommand\ptveto{p_{\sss T}^{(veto)}}
\newcommand\ptll{p_{\sss T}^{(ll)}}
\newcommand\ptmin{p_{\sss T}^{(min)}}

\newcommand \ket [1] {|{#1}\rangle}
\newcommand \bra [1] {\langle  {#1}|}
\newcommand \slsh [1] {\not\!{#1}}

\newcommand\gw{g_{\sss W}}
\newcommand\muf{\mu_{\sss F}}
\newcommand\mur{\mu_{\sss R}}

\preprint{
 CERN-TH/2008-104;  ITP-UU-08/27\hfill\\
 NIKHEF-2008-005;  ITFA-2008-16\hfill\\
 Cavendish-HEP-08/06
 }
\title{Single-top hadroproduction in association with a $W$ boson}
\author{Stefano Frixione%
  \thanks{On leave of absence from INFN, Sez. di Genova, Italy}\\
  PH Department, TH Unit, CERN, CH-1211 Geneva 23, Switzerland\\
  ITPP, EPFL, CH-1015 Lausanne, Switzerland\\
  E-mail: \email{Stefano.Frixione@cern.ch}}
\author{Eric Laenen\\
  ITFA, University of Amsterdam,
  Valckenierstraat 65, 1018 XE Amsterdam, \\
  Nikhef Theory Group,  Kruislaan 409, 1098 SJ Amsterdam, The Netherlands\\
  ITF, Utrecht University, Leuvenlaan 4, 3584 CE Utrecht\\
  E-mail: \email{Eric.Laenen@nikhef.nl}}
\author{Patrick Motylinski\\
  Nikhef Theory Group,  Kruislaan 409, 1098 SJ Amsterdam, The Netherlands\\
  E-mail: \email{patrickm@nikhef.nl}}
\author{Bryan R.\ Webber\\
  Cavendish Laboratory, 
  J.J. Thomson Avenue, Cambridge CB3 0HE, U.K.\\
  E-mail: \email{webber@hep.phy.cam.ac.uk}}
\author{Chris White\\
  Nikhef Theory Group,  Kruislaan 409, 1098 SJ Amsterdam, The Netherlands\\
  E-mail: \email{cwhite@nikhef.nl}}
\abstract{We present the calculation of the $Wt$ single-top production 
channel to next-to-leading order in QCD, interfaced with 
parton showers within the MC@NLO formalism. This channel provides a 
complementary way of investigating the properties of the $Wtb$ vertex, 
with respect to the $s$- and $t$-channels.
We pay special attention to the separation of this process from top quark 
pair production.}

\begin{document}

\section{Introduction}
\label{sec:introduction}
The top quark was discovered in 1995 and has been intensively studied at the
Tevatron. It will also be a prime object of study at the forthcoming 
Large Hadron Collider (LHC). 
With production rates far in excess of current experiments,
the LHC can be thought of as a veritable top quark factory. The large mass of
the quark, near the electroweak scale, enables detailed scrutiny of its
interactions unshrouded by hadronisation effects (which are suppressed by
powers of the perturbative momentum scale). Furthermore, effects of physics
beyond the Standard Model are thought to lie around the electroweak scale,
thus the top quark sector provides a valuable window for new physics.

A useful process to study in this regard is the production of single top
quarks via the weak interaction, as it offers a relatively clean probe of the
properties of the heavy quark. Although there was recently at least a
3$\sigma$ evidence of single top production at both Tevatron experiments (see
e.g. refs.~\cite{RW,Abazov:2008kt}), the cross section is sufficiently
small (within the SM) as to preclude detailed scrutiny. However, one 
expects a significant number of single top events at the LHC, where 
the centre of mass energy is much higher.

Accurate estimates of rates and kinematic distributions at hadron colliders
necessitate higher-order perturbative computations in QCD.
Furthermore, in order to optimise acceptance cuts in experimental
analyses or perform full detector simulations, one needs realistic
hadron-level events. These are obtained using Monte Carlo event generators
that incorporate the simulation of parton showers and hadronisation
models. The complementary benefits of fixed-order computations and parton
shower simulations are by now well-known, as are the advantages of combining
them into a framework which utilizes the benefits of each of them. The MC@NLO
approach~\cite{Frixione:2002ik,Frixione:2003ei} provides a way of achieving
this, by allowing one to match cross sections computed at NLO in QCD with an
event generator. No modifications to the latter are necessary, and colour
coherence is preserved by the matching procedure. Therefore, existing 
parton shower Monte Carlos with no add-ons can be used for this purpose.

There are three distinct production modes for single top quarks.
In earlier work~\cite{Frixione:2005vw} we included
two of these modes, the $s$- and $t$-channel processes, into the MC@NLO
framework. In this paper, we implement the production of a single top quark in 
association with a final state $W$ boson, thereby completing the description 
of single top processes in hadronic collisions at this level of accuracy, 
including spin correlations of decay products, which we incorporate
as explained in ref.~\cite{Frixione:2007zp}. 

As is well known, the inclusion of higher order corrections for the
$Wt$ channel has challenging peculiarities due to interference 
with the $t\bt$ process. This interference becomes extremely
large in certain phase-space regions, and apparently 
renders the perturbative computation of the $Wt$ cross section meaningless. 
Nevertheless, owing to distinct features of $t\bt$ production, several
{\em definitions} of the $Wt$ channel have been given in the literature,
each with the aim of recovering a well-behaved expansion in $\as$. The problem
of interference in fact affects any computation that considers contributions
beyond the leading order, i.e. at least ${\cal O}(\gw^2\as^2)$. The
cross section at this order has been previously presented in
refs.~\cite{Tait:1999cf,Belyaev:1998dn,Kersevan:2006fq}, where only
tree-level graphs were considered, and in 
refs.~\cite{Zhu:2002uj,Campbell:2005bb,Cao:2008af}, where one-loop
contributions were included as well. In ref.~\cite{Beccaria:2007tc},
the calculation of NLO electroweak effects has been carried out, using
the results of ref.~\cite{Campbell:2005bb} as a basis.
We shall comment on some of these papers in the following.

The aim of this paper is to critically examine the definition of the
$Wt$ channel, and to propose two options that can be used in the
context of an NLO calculation interfaced with parton showers.
Although these two options are well defined even in the absence of 
kinematic cuts, we will in addition consider final-state cuts that will
further help the separation of the $Wt$ and $t\bt$ processes.
Such cuts, at variance with those of previous approaches 
(except ref.~\cite{Kersevan:2006fq}), are
easily applied in an experimental environment. The implementation
of two definitions of the $Wt$ channel in the same framework will
allow us to estimate the theoretical systematics potentially affecting
the extraction of the $Wt$ signal from data. Although the implementation 
of any definition of the $Wt$ channel into any higher-order computation 
is technically non-trivial (and MC@NLO is no exception to that),
we argue and make it plausible that an NLO+parton shower framework is
uniquely suited to the discussion of the physical consequences of such a
definition.

The structure of the paper is as follows. In section~\ref{sec:prob} we 
give a first general discussion of the problem of interference with $t\bt$ 
production. In section~\ref{Wtprod} we recall
the various modes of single top quark production, before giving
some technical details on the implementation of $Wt$ production
in MC@NLO, and on the underlying NLO computation. In section~\ref{inter} 
we analyse the problem of interference in more detail, discuss the approaches 
in the literature, and propose two definitions of the $Wt$ channel that
can be used in a parton shower context and in an experimental
analysis. We proceed in section~\ref{results}
to present the results obtained by applying these two definitions
in MC@NLO simulations. We discuss physical implications and conclude in 
section~\ref{discussion}. Certain technical details are collected 
in the appendices.

\section{Nature of the problem}
\label{sec:prob}
In the perturbative computation of $Wt$ production, one must consider
all possible partonic processes with final states
\beq
t+W+\sum_i X_i\,.
\eeq
Here, $\{X_i\}$ is a set of particles (partons in a QCD 
computation), whose multiplicity increases as the perturbative order 
increases. At the leading order (LO) in the SM, ${\cal O}(\gw^2\as)$,
such a set is empty, and the underlying partonic process is
\beq
bg\,\longrightarrow\, tW\,.
\label{eq:LO}
\eeq
When next-to-leading order (NLO) corrections in $\as$ are considered, 
contributions (e.g. $gg\to tW\bb$) will appear such that
\beq
\{X_i\}\equiv \bb\,.
\eeq
Some of the relevant Feynman diagrams will feature the $W\bb$ pair
originating from a $\bt$ internal line; in other words, the momentum
flowing in the $\bt$ propagator will be \mbox{$k_W+k_{\bb}$}. Therefore,
in the computation of certain observables, one will need to integrate
over the region
\beq
M_{\bb W}^2\equiv (k_W+k_{\bb})^2\simeq m_t^2\,.
\label{eq:toppeak}
\eeq
When this is the case, a divergence is encountered, which is regulated
only by using a finite width for the (anti-)top, $\Gamma_t\ne 0$. However,
a non-zero top width is an all-order result in $\gw$. Thus, the inclusion
of higher-order QCD corrections forces one to include electroweak
corrections to all orders so as to avoid divergences, and this
potentially spoils the power counting in $\gw$, according to which
eq.~(\ref{eq:LO}) is the LO contribution to $tW$ production.

It turns out that indeed the region of eq.~(\ref{eq:toppeak}) causes
severe problems, and as a result NLO QCD corrections are much larger
than the LO result obtained with eq.~(\ref{eq:LO}).
One possible way out is that of considering only
observables which are exclusive enough to allow one to exclude, 
through final-state cuts, the resonance region eq.~(\ref{eq:toppeak}).
This is easily done in a fixed-order, parton-level theoretical
computation, where the $W$ and the $\bb$ are easily accessible.
It becomes however a more indirect procedure in the context of a 
parton shower simulation,
and impossible in a real experiment. Note also that, in order to avoid
biases in an MC simulation, very loose cuts or no cuts at all will have
to be imposed in the computation of underlying matrix elements, with the
result that the majority of the events generated, being close to 
the $\bt$ resonance, will actually be thrown away by final-state cuts, 
leading to a very low efficiency.

The dominance of the $\bt$ resonance suggests that a possible approach
is that of simply considering $W^+W^- b\bb$ final states (with the
possibility of also including production spin correlations as well, for
the fermions resulting from the decays of the $W$'s -- see e.g.
ref.~\cite{Kauer:2001sp} for di-leptonic decays). In this way, 
what has previously been denoted as the LO contribution to $Wt$ production,
eq.~(\ref{eq:LO}), can be seen as a correction to $W^+W^- b$ observables
(i.e., observables inclusive in the final-state $\bb$ present in the 
$W^+W^- b\bb$ matrix elements). The way in which this contribution 
is taken into account is a matter of careful definition, which is
especially delicate (because of the double counting problem) when 
interfacing the matrix elements to parton showers --
see what was done in ref.~\cite{Kersevan:2006fq}.

By emphasising the role of $W^+W^- b(\bb)$ final states, these
approaches have an immediate connection with data, and finding
the $Wt$ ``signal'' is thus a matter of a careful counting experiment.
The key question is: are these predictions accurate enough for 
the counting to be reliable? Indeed, in these approaches the 
problem of full NLO corrections, whose knowledge is known to be 
crucial in top physics, is not considered.

The computation of NLO corrections to $Wt$ production in QCD can only 
be seriously undertaken by
recovering a meaningful definition of a perturbative expansion
whose LO contribution is that of eq.~(\ref{eq:LO}). We have seen
that the problem arises because of the interplay between $\as$
and $\gw$ expansions, and specifically because of the necessity
of considering all-order contributions in $\gw$.
The key observation is that the all-order result in $\gw$ that we
need is associated with a {\em decay}, whereas we are only interested in
the role of EW interactions in the {\em production} process.
A meaningful expansion in $\as$ of the $Wt$ cross
section could therefore be achieved if electroweak effects in
production and decay could be disentangled. Such a
procedure can never be fully consistent theoretically, but it
can be given an operational meaning, which can be tested 
experimentally. The rest of the paper will be devoted to 
separating production and decay EW effects in the context of MC@NLO.

\section{$Wt$ production at NLO}
\label{Wtprod}
In this section we introduce the various single top production modes at LO,
before giving some details of the calculation of the $Wt$ channel at NLO in
QCD. Technical details regarding the implementation in MC@NLO are presented,
although as these are similar to those relevant to the other single top
production channels~\cite{Frixione:2005vw} we refer the reader to
previous publications where appropriate. Since in this section we
will be mainly concerned with the computation of the matrix elements
entering the NLO predictions, and with the construction of the MC subtraction
terms for MC@NLO, we can safely ignore the issues discussed
in sect.~\ref{sec:prob}, but will return to them in sect.~\ref{inter}.

\subsection{Born level}
There are three modes for the production of a single top quark at 
LO\footnote{We assume throughout the paper that a 5
 flavour scheme can be used for the quark sector, where the $b$ quark
 is included in the initial state parton distributions. In principle
 the calculation could also be formulated in a 4 flavour scheme where all
 $b$ quarks are generated in the final state.}.
\begin{figure}
  \begin{center}
      \epsfig{figure=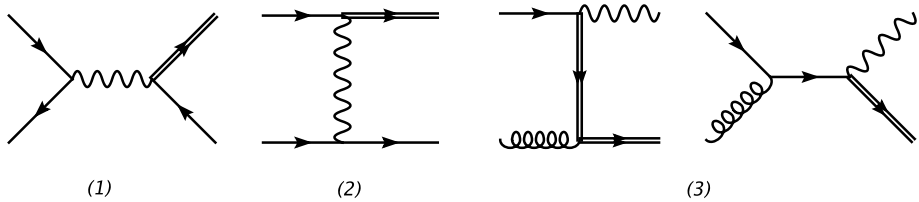,width=0.8\textwidth}
\caption{
Leading order diagrams for single-$t$ production in the 
(1) $s$-channel, (2) $t$-channel and (3) $Wt$-channel. The $t$-quark
 line is doubled.}
\label{modes} 
  \end{center}
\end{figure}
Each of these provides a separate and complementary means of studying the
$Wtb$ vertex: the $s$-channel mode (fig.~\ref{modes}(1)) involves a time-like,
off-shell vector boson which may reveal new resonances; the $t$-channel mode
(fig.~\ref{modes}(2)) involves a spacelike boson and is sensitive to flavour
changing neutral currents (FCNC's). Finally, the $Wt$ mode
(fig.\ref{modes}(3)) has an on-shell $W$ boson, and thus provides a
complementary source of information on the flavour structure of the vertex
(i.e. $V_{tb}$), and its chiral properties. Based on expectations from
the SM, this latter process has far too small a cross section to be observed
at the Tevatron, but is expected to be significant at the LHC.

As shown in fig.~\ref{modes}(3), the lowest order partonic process for
$Wt$ production is
\begin{equation}
  \label{eq:1}
  b(p_1) + g(p_2) \longrightarrow t(k_1) + W^-(k_2)\,.
\end{equation}
Our description is confined to $W^-t$ production. As explained in appendix
\ref{sec:wbart-prod-same}, for $W^+\bar{t}$ production no new diagrams
need be computed, so that the description below suffices.
Where no confusion is possible in the following, we denote the light quark
attached to the $Wt$ vertex as a $b$, implicitly representing any CKM-allowed
down-type quark. 

It is convenient to introduce the following invariants
\begin{equation}
  \label{eq:p2}
  s = (p_1+p_2)^2, \quad t_1 = t-m_t^2 = (k_1-p_1)^2-m_t^2, \quad
u_1 = u-m_t^2 = (k_2-p_1)^2-m_t^2
\end{equation}
such that $s+t_1+u_1 = m_W^2-m_t^2$, with $m_t$ denoting the top quark mass.
We treat all other quarks as massless, including the $b$ quark. 
The lowest order cross section can be written
\beqn
  \label{eq:4}
d\sigma^{(0)}&=&{\cal M}^{(0)}d\phi_2\,,
\\*
{\cal M}^{(0)} &=& \frac{1}{2s}\frac{1}{4}\frac{1}{N(N^2-1)} \gs^2
\frac{\gw^2}{8}\, |{\cal A}^{(0)}|^2 \, 
\eeqn
where $\gs$, $\gw$ are the QCD and EW coupling constants; $N$ the number of
colours; $d\phi_2$ denotes the two-body final state phase space, and the
spin-summed Born-level matrix element is given by
\begin{multline}
  \label{eq:3}
|{\cal A}^{(0)}|^2 = 16 N C_F \Bigg\{ -\left(\frac{s}{u_1}+\frac{u_1}{s}
  \right)\left(1+\frac{m_t^2}{2m_W^2} \right) 
\\* 
+2\,\frac{m_W^2 (s+u_1)u_1+m_t^2 m_W^2 s}{s u_1^2} 
  \left(1-\frac{m_t^2}{2m_W^2}-\frac{m_t^4}{2m_W^4} \right)  
\\*
-\frac{m_W^4}{s u_1}\left(2-\frac{3m_t^2}{m_W^2}+\frac{m_t^6}{m_W^6} \right) -
\frac{m_t^2}{m_W^2} \Bigg\} \,.
\end{multline}

\subsection{NLO computation}
In order to implement a process in MC@NLO, one must cancel the singularities
arising in the real and virtual graphs in the particular subtraction formalism 
of refs.~\cite{Frixione:1995ms,Frixione:1997np} and denoted FKS henceforth. 
In the FKS formalism, the NLO cross section is expressed in terms 
of the finite quantities that result from the cancellation of the
(universal and process-independent) soft and collinear poles arising
from virtual and real corrections. In order to obtain these finite quantities 
for a given process, one needs the real matrix
elements computed in $4$ dimensions, and the virtual corrections
computed in $d$ dimensions. Although not strictly necessary in the FKS
method, a $d$-dimensional virtual computation was carried out
to fix the convention for its finite part.

The real and virtual matrix elements relevant to $Wt$ production
can be extracted from the NLO calculation of the $Wc$ 
process~\cite{Giele:1996kr}, for which the diagrams are almost
identical\footnote{Essentially, only the $W+(c\bar{c})$ channel 
has no analogue in the present case.}.
Given that the computation of ref.~\cite{Giele:1996kr} adopted the
phase-space slicing method~\cite{Giele:1992vf,Giele:1993dj,Keller:1998tf} 
for cancelling the real and virtual singularities, we converted the 
conventions of the phase-space slicing into those of FKS. In order to 
check that this was carried out correctly, we repeated the calculation 
of the matrix elements.

\subsubsection{Virtual corrections}
\label{sec:virt}
As stated above, we calculated the one-loop virtual diagrams 
in dimensional regularisation in $d=4-2\epsilon$ dimensions.
Tensor and vector integrals were reduced to scalar integrals 
using the standard Passarino-Veltman algorithm~\cite{Passarino:1978jh}. 
For generation and evaluation
of the relevant amplitudes, we utilised the Mathematica packages
FeynArts~\cite{Hahn:2000kx} and FeynCalc~\cite{Mertig:1990an}, together with
the FORM~\cite{Vermaseren:2000nd} computer program. As discussed previously,
the results were checked against the virtual contributions obtained for 
$Wc$ production in ref.~\cite{Giele:1996kr}, and found to be in agreement
after the necessary analytical continuation (from $m_c<m_W$ to $m_t>m_W$).
We have also checked, analytically and numerically, that our results 
for the scalar integrals are 
in agreement with those of ref.~\cite{Ellis:2007qk}.

The virtual corrections are a Laurent series in the parameter $\epsilon$, with
double and single poles arising from soft, initial-state collinear,
and ultraviolet (UV) singularities\footnote{Final-state collinear
singularities are absent due to regularisation by the top mass.}. 
To remove UV poles we renormalise the top
quark mass using an on-shell condition, with the QCD coupling in the
$\overline{\rm MS}$ scheme modified such that the top quark loop contribution
is subtracted on-shell. This particular scheme~\cite{Collins:1978wz} ensures
that the top quark virtual contributions decouple in the limit of small
external momenta. Specifically, the coupling is renormalised through
next-to-leading order as
\begin{equation}
  \label{eq:5}
  \gs \rightarrow \gs(\mur^2) \left[1 +
\frac{\alpha_s(\mur^2)}{8\pi}\left(\frac{-1}{\epsilon}+\gamma_E
-\ln 4\pi\right) \left(\frac{\muf^2}{\mur^2} \right)^\epsilon \beta_0 +
\frac{\alpha_s(\mur^2)}{8\pi}\, \frac{2}{3}
\ln\left(\frac{\mur^2}{m_t^2}\right) \right]
\end{equation}
where $\muf$ is the factorisation scale, and $\mur$ the renormalisation
scale. Furthermore, \mbox{$\beta_0 = (11 C_A - 2 n_f)/3$} with $n_f$ 
equal to the number of light flavors (here five) plus one.  From this 
condition one can derive the following relation for the renormalised 
QCD coupling
\begin{equation}
  \label{eq:6}
  \mur^2 \frac{d\gs(\mur^2)}{d\mur^2} = - \gs(\mur^2)
\frac{\alpha_s(\mur^2)}{8\pi}\left(\beta_0+\frac{2}{3} \right)
+ {\cal O}(\gs^5)\,,
\end{equation}
which indeed removes the top quark loop from the $\beta$-function, such
that the number of light flavours implemented in the code is $n_f$. The
renormalisation of the top quark mass is given by
\begin{multline}
  \label{eq:7}
  m_t \rightarrow m_t + \delta m_t = \\ m_t \left[1 +
 \frac{\as(\mur^2)}{4\pi}\,
C_F\left(\frac{-3}{\epsilon}+3\gamma_E -3\ln 4\pi -4 - 3\ln
 \left(\frac{\mur^2}{m_t^2}\right)\right) \right]\,.
\end{multline}
In contrast to the $s$ and $t$ single top production channels, the top quark
occurs as an internal line in the Born amplitude. This gives rise to a
contribution to the amplitude coming from the expansion of the renormalised
top quark propagator
\begin{equation}
  \label{eq:12}
  \frac{i}{\slsh{p}-(m_t+\delta m_t)} = \frac{i}{\slsh{p}-m_t} +
  \frac{i}{\slsh{p}-m_t} \delta m_t \frac{1}{\slsh{p}-m_t}.
\end{equation}
The first term in this expansion gives the Born amplitude ${\cal A}^{0}$,
whereas the second gives a modified amplitude ${\cal A'}^{(1)}$ involving the
``squared'' top quark propagator. In terms of the finite remainder of 
$\delta m_t$, the cross section receives the contribution
\begin{equation}
  \label{eq:15}
  d\sigma^{(1,V)}_{\delta m_t} = (\delta m_t)_{\mathrm{finite}}
\frac{1}{2s}\frac{1}{4}\frac{1}{N(N^2-1)} \gs^2
\frac{\gw^2}{8}\, 
\left({\cal A}^{(0)}{\cal A'}^{\dag(1)} + 
{\cal A'}^{(1)}{\cal A}^{\dag(0)}\right)
\, d\phi_2\,,
\end{equation}
where
\begin{multline}
{\cal A}^{(0)}{\cal A'}^{\dag(1)} + 
{\cal A'}^{(1)}{\cal A}^{\dag(0)} = 
-16 N C_F \frac{m_t^2}{m_W^2 s u_1^3}\Bigg\{ \left[4m_t^6 s +
4m_t^4 s(m_W^2+u_1)\right] 
\\* 
+m_t^2\left[-8m_W^4 s + 2m_W^2su_1 +
u_1(u_1+t_1)(s+2t_1) \right] 
\\* +u_1 \left[u_1 t_1
(u_1+t_1)+m_W^2(-s(u_1-2t_1)+4 t_1(u_1+t_1)) \right] \Bigg\}\,.
\end{multline}
Following the renormalisation procedure, soft and collinear singularites
remain and are proportional to the Born cross section such
that the virtual terms may be written (c.f. eqs.~(3.1) and~(3.2) in
ref.~\cite{Frixione:1995ms})
\begin{multline}
  \label{eq:8}
  d\sigma^{(1,V)} =C_\epsilon \gs^2 \, \left[-\frac{2}{\epsilon^2}(C_F+C_A) +
\frac{2}{\epsilon}C_A\left(\ln\frac{s}{m_t^2}+\ln\frac{-u_1}{m_t^2}-
\ln\frac{-t_1}{m_t^2}\right)\right.
\\*
\left. +\frac{1}{\epsilon}C_F\left(4\ln\frac{-t_1}{m_t^2}-5\right) 
-\frac{1}{\epsilon}\beta_0\right] d\sigma^{(0)}_{4-2\epsilon} +
d\sigma^{(1,V)}_{\mathrm{finite}},
\end{multline}
where $d\sigma^{(0)}_{4-2\epsilon}$ is the Born cross section in
$4-2\epsilon$ dimensions, and
\begin{equation}
  \label{eq:13}
  C_\epsilon = \frac{(4\pi e^{-\gamma_E})^\epsilon}{16\pi^2}
  \left(\frac{\muf^2}{m_t^2} \right)^\epsilon \,.
\end{equation}
These remaining poles are cancelled
by similar singularities in the real contributions and by the collinear
counterterms that arise from the renormalisation of the parton
densities, in a form prescribed by the FKS formalism.

\subsubsection{Real corrections}
Since only initial-state collinear singularities are present in
the process considered here, it is not necessary to partition the
phase-space according to the FKS prescription, as done in 
ref.~\cite{Frixione:2005vw} for the case of $s$- and $t$-channel
single-top production (see sect.~2.1.2 of that paper).

Therefore, the only technical difficulty in assembling the real
corrections is in the explicit calculation of the diagrams\footnote{There
is a subtlety involving the construction of the local initial-state
collinear counterterms used in numerical codes, as in all cases in 
which a gluon is exchanged. This issue is discussed in 
appendix~\ref{sec:calc-tild-helic}.}.
As explain previously, we used the results of the $Wc$ calculation
presented in ref.~\cite{Giele:1996kr}. There, the matrix elements
were computed by considering the fictitious $W$ boson
decays
\begin{equation}
  \label{eq:9}
  W(q)\;\longrightarrow\;Q(p)+\bb_3(q_3)+g_4(q_4)+g_5(q_5)\,
\end{equation}
and
\begin{equation}
  \label{eq:10}
  W(q)\;\longrightarrow\;Q(p)+\bb_3(q_3)+b_4(q_4)+\bb_5(q_5)\,,
\end{equation}
where $Q$ denotes generically the only quark with mass different from zero.
These were decomposed into colour-ordered amplitudes, and computed using 
FORM~\cite{Vermaseren:2000nd}. The $Wc$ and $Wt$ matrix elements were
subsequently obtained by crossing and summing over colour orders. This saved
somewhat on computational effort, since the colour-ordered amplitudes served
a second purpose, namely in the accounting of colour connections in the parton
shower stage of the MC@NLO construction. 

As a further check on the real computation, we compared the results we
obtained from our calculation against the corresponding tree level matrix
elements generated using the \texttt{MADGRAPH} 
program~\cite{Maltoni:2002qb,Alwall:2007st}. 
We found agreement in all cases. 

\subsection{Implementation in MC@NLO}
Several processes have by now been computed in the MC@NLO formalism, and we 
will therefore refrain from describing here the necessary steps for
the implementation of a new reaction. The interested reader can find
all details in refs.~\cite{Frixione:2002ik,Frixione:2003ei,Frixione:2005vw},
which also report all the relevant analytic formulae we need (these
are process-independent, and therefore no new computation is specifically 
required for $Wt$ production). In this section, we will limit
ourselves to giving the necessary information for the construction
of the MC subtraction terms (which are process dependent) necessary
for the matching of the NLO computation with 
HERWIG~\cite{Corcella:2000bw,Corcella:2002jc}.

One may write the MC subtraction terms as (see 
refs.~\cite{Frixione:2003ei,Frixione:2005vw})
\beq
d\sigma\xMCB=\sum_{i}\sum_{L}\sum_{l}
d\sigma_{i}^{(L,l)}\xMCB\,,
\label{MCatas}
\eeq 
where $i$ sums over the different partonic subprocesses, and 
$L\in\{+,-,f_1\}$ labels the parton leg from which the FKS parton is emitted
(respectively: incoming parton along the $+z$ direction; incoming parton along
the $-z$ direction; final state top quark). The index $l$ runs over the
different colour structures. If $q_\alpha$ and $q_\beta$ denote the 4-momenta 
of the colour partners relevant to a given branching, the shower scale
associated with the branching is 
\beq 
E_0^2=|q_\alpha\mydot q_\beta|.
\label{shwrscale}
\eeq 
The individual subtraction terms on the right-hand side of
eq.~(\ref{MCatas}) have the form (see eqs.~(5.2)--(5.5) in
ref.~\cite{Frixione:2003ei})
\beqn d\sigma_{i}^{(+,l)}\xMCB &=&
\frac{1}{z_+^{(l)}} f_a^{\Hone}(\bx_{1i}/z_+^{(l)})f_b^{\Htwo}(\bx_{2i})\,
d\hat{\sigma}_{i}^{(+,l)}\xMCB d\bx_{1i}\,d\bx_{2i}\,,
\label{eq:spl}
\\
d\sigma_{i}^{(-,l)}\xMCB &=&
\frac{1}{z_-^{(l)}}
f_a^{\Hone}(\bx_{1i})f_b^{\Htwo}(\bx_{2i}/z_-^{(l)})\,
d\hat{\sigma}_{i}^{(-,l)}\xMCB d\bx_{1i}\,d\bx_{2i}\,,
\label{eq:smn}
\\
d\sigma_{i}^{(\fo,l)}\xMCB &=&
f_a^{\Hone}(\bx_{1f})f_b^{\Htwo}(\bx_{2f})\,
d\hat{\sigma}_{i}^{(\fo,l)}\xMCB d\bx_{1f}\,d\bx_{2f}\,,
\label{eq:sfo}
\eeqn 
where $f_{a,b}^{(H_{1,2})}$ are the initial state parton distributions, 
and $z_L$ labels 
the partonic momentum fraction carried by the FKS parton. The short-distance 
cross sections $d\hat{\sigma}_{i}^{(L,l)}$ can be obtained from 
eqs.~(5.6)--(5.8) of ref.~\cite{Frixione:2003ei}
\beqn
d\hat\sigma_{i}^{(\pm,l)}\xMCB &=& \frac{1}{{\cal N}}\frac{\as}{2\pi}\,
\frac{d\xi_\pm^{(l)}}{\xi_\pm^{(l)}}dz_\pm^{(l)}
P_{a^\prime b^\prime}(z_\pm^{(l)})\,
\dsb_{i^\prime}
\stepf\left((z_\pm^{(l)})^2-\xi_\pm^{(l)}\right);
\label{eq:shin}
\\
d\hat\sigma_{i}^{(\fo,l)}\xMCB &=& \frac{\as}{2\pi}\,
\frac{d\xi_{\fo}^{(l)}}{\xi_{\fo}^{(l)}}dz_{\fo}^{(l)}
P_{gq}(z_{\fo}^{(l)})\,\dsb_{i^\prime}
\stepf\left(1-\xi_{\fo}^{(l)}\right)
\stepf\left(z_{\fo}^{(l)}-
\frac{m_t}{E_0\sqrt{\xi_{\fo}^{(l)}}}\right),\phantom{aa}
\label{eq:shout}
\eeqn 
where ${\cal N}=1$ for quark branchings, and 2 for gluon branchings
(since the gluon has two colour partners); see 
refs.~\cite{Frixione:2003ei,Frixione:2005vw} for the definition
of $\dsb$. The various partonic subprocesses
contributing to the $Wt$ channel are listed in table~\ref{tab:Wtch}
(the initial states $bq$ and $qb$ represent also $b\bq$ and $\bq b$).
\begin{table}
\begin{center}
\begin{tabular}{|c||c|c|}
\hline
NLO & 
  $(b,g;t,W)$ & $(g,b;t,W)$ \\\hline\hline
$(g,g;t,W,\bb)$ &
  $+(\bp_1\mydot \bp_2)$ &
  $-(\bp_1\mydot \bp_2)$ \\\hline\hline
$(b,q;t,W,q)$   & 
  $-(\bp_1\mydot \bp_2,\bp_2\mydot \bk_1)$ & \\\hline
$(q,b;t,W,q)$   & & 
  $+(\bp_1\mydot \bp_2,\bp_1\mydot \bk_1)$ \\\hline
$(b,\bb;t,W,\bb)$ &
  $-(\bp_1\mydot \bp_2,\bp_2\mydot \bk_1)$ & \\\hline
$(\bb,b;t,W,\bb)$ & & 
  $+(\bp_1\mydot \bp_2,\bp_1\mydot \bk_1)$ \\\hline
$(b,b;t,W,b)$     & 
  $-(\bp_1\mydot \bp_2,\bp_2\mydot \bk_1)$ & 
  $+(\bp_1\mydot \bp_2,\bp_1\mydot \bk_1)$ 
  \\\hline\hline
$(b,g;t,W,g)$ & 
  $+,\fo(\bp_1\mydot \bk_1)$; 
  $-(\bp_1\mydot \bp_2,\bp_2\mydot \bk_1)$ & \\\hline
$(g,b;t,W,g)$ & &
  $-,\fo(\bp_2\mydot \bk_1)$;
  $+(\bp_1\mydot \bp_2,\bp_1\mydot \bk_1)$ \\\hline
\end{tabular}
\end{center}
\caption{\label{tab:Wtch}
Short-distance contributions to MC subtraction terms. The two columns 
correspond to the two possible Born cross sections. For a given process, 
the entries show the emitting legs, and in round brackets the value(s) of 
the shower scale(s) $E_0^2$ (up to a sign).
}
\end{table}
The shower scales to 
be used in eqs.~(\ref{eq:shin}) and~(\ref{eq:shout})
are equal to the absolute values of the dot products given in 
table~\ref{tab:Wtch}. From table~\ref{tab:Wtch}, we see that ${\cal N}=2$ 
for all branchings in the processes with $\{b,q\}$, $\{b,\bb\}$ and
$\{b,b\}$ initial states, and for branchings from leg $-$ and leg $+$
in the processes $(b,g;t,W,g)$ and $(g,b;t,W,g)$ respectively.

When interfacing the NLO calculation with a parton shower, 
it is also necessary to supply a colour flow
to the Monte Carlo. When a given subprocess can have more than one colour
flow, the one that is given to the parton shower is chosen on a statistical
basis. The weights that govern these probabilities are formed using leading
order (in the number of colours $N$) approximations to the squared matrix
elements of the corresponding tree-level diagrams. In our case, these have
already been calculated, as the real emission contributions were computed
using colour ordered amplitudes. Given that there is a one-to-one mapping
between the colour ordered amplitudes and the various colour flows, 
one can choose to use these colour ordered amplitudes in the statistical 
determination of the colour flow\footnote{Other choices are possible, as 
one is free to modify the above procedure by terms subleading in the number 
of colours $N$.}.

\section{Interference between $t\bar{t}$ and $Wt$ production}
\label{inter}
In section~\ref{Wtprod} we have discussed the calculation of the $Wt$
cross section within the FKS subtraction formalism, and its subsequent
implementation in MC@NLO. However, the use of such a Monte Carlo event
generator assumes that the $Wt$ channel is well-defined beyond
LO. In fact, as already outlined in sect.~\ref{sec:prob},
this is not the case, and the theoretical definition of this production 
channel is not straightforward.

\begin{figure}
  \begin{center}
      \epsfig{figure=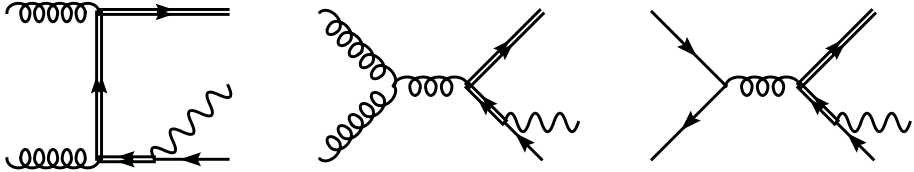,width=0.8\textwidth}
\caption{Diagrams that are doubly-resonant, in the sense that 
the intermediate $\bar{t}$ can be on-shell.}
\label{fig:dr} 
  \end{center}
\end{figure}

At LO, the $Wt$ pair is produced through the reaction of 
eq.~(\ref{eq:1}), and its cross section is smaller
than that of top pair production by a factor of about 15. 
Beyond LO, some of the Feynman graphs that contribute to the $Wt$ 
channel and that are dominant in the region of eq.~(\ref{eq:toppeak})
are shown in fig.~\ref{fig:dr}.
These diagrams can be interpreted as the production of a $t\bt$ pair at LO, 
with subsequent decay of the $\bt$ into a $\bb W$ pair. 
It follows that the set of Feynman graphs contributing to
$gg\to tW\bb$ or $q\bq\to tW\bb$ can be divided into two subsets,
that are customarily denoted as {\em doubly resonant} and
{\em singly resonant}, the former being those depicted
in fig.~\ref{fig:dr} or analogous ones.
The problem discussed in sect.~\ref{sec:prob}
can be formulated in terms of the interference between these two 
subsets of graphs, and physically
interpreted as the interference between $Wt$ and $t\bt$ production.
This interpretation is consistent with our aim of emphasising the
role of electroweak interactions in the production mechanism
of the $t$ and of the $W$.
The interference would not cause any problem if the contribution 
from the doubly-resonant diagrams was small. However, a large 
increase of the cross section occurs when the $\bt$ propagator 
becomes resonant i.e. in the kinematic region of eq.~(\ref{eq:toppeak}).
It follows that the na\"{\i}ve calculation of the $Wt$ channel,
defined as that resulting from the straightforward computation
of Feynman diagram as discussed in sect.~\ref{Wtprod},
becomes meaningless in perturbation theory.
In this section, we will discuss how it is possible to separate out 
$Wt$ and $t\bt$ production, and recover a workable
definition of the $Wt$ channel.

\subsection{Previous approaches}
\label{sect:NLOcomp}
This interference problem has already been discussed in several papers.
Given that the problem here only arises for $gg$ and $q\bq$ initial states,
present in real-emission processes, it is not peculiar to full higher-order 
computations, but also occurs in those that only include tree-level graphs.
Each of the previous $Wt$ computations beyond LO has introduced a strategy
for isolating the $Wt$ production channel. We briefly discuss each of
them in turn.

In ref.~\cite{Belyaev:1998dn}, the suggestion was made to place a 
cut on the invariant mass $M_{\bb W}$ of the $W\bar{b}$ pair, which can be
written in the form:
\begin{equation}
|M_{\bb W}-m_t|>\kappa\Gamma_t,
\label{mbWcut}
\end{equation}
where $\Gamma_t$ is the width of the top quark. This cut thus removes events
from the region of phase space corresponding to the $\bar{t}$ resonance. It
efficiently reduces the $t\bt$ interference to the $Wt$ channel, but cannot 
be directly applied in an event generation/experimental situation, where one
is unable to unambiguously identify the $b$ quark and $W$ boson which 
originate in the hard interaction.

In ref.~\cite{Tait:1999cf}, the $Wt$ signal is defined by subtracting, 
at the level of squared amplitudes, the $t\bt$ cross section multiplied by
$t\to Wb$ branching ratio; the procedure is defined in a fully-inclusive
way. A comparison was made with the invariant mass cut approach 
of ref.~\cite{Belyaev:1998dn}, and the resulting cross section found 
to agree if the choice $\kappa\sim 15$ was made. 

As discussed in sect.~\ref{sec:prob}, the approach of 
ref.~\cite{Kersevan:2006fq} addresses the problem in a more general
context, i.e. that of simulating $W^+W^- b(\bb)$ production in a
realistic parton-shower environment. As such, the definition of the
$Wt$ channel does not need to be directly considered, except for 
the necessity of avoiding double counting. This way of by-passing
the interference problem is correct, and facilitates the
comparison with data. However, it does not allow one to include
the full NLO corrections into the results, because of both 
problems of practice (the relevant one-loop, six-point amplitudes -- 
or eight-point amplitudes, if fully decayed $W$'s are considered --
are not available yet), and of principle (how to combine
the one-loop contribution for $W^+W^- b$ with the rest).

The full NLO computation of ref.~\cite{Zhu:2002uj} follows the same
type of strategy as ref.~\cite{Tait:1999cf}, where the total cross section
is modified by a subtraction term which effectively removes the 
$t\bt$-like contribution. However, few details on how this subtraction 
term is defined are given. The results are presented only for total rates.

Finally, a definition of the $Wt$ channel within a fully
differential NLO computation was presented in ref.~\cite{Campbell:2005bb}.
There, the suggestion was made to place a cut on the transverse momentum 
$\pt^b$ of the ``additional $b$ quark that appears at the next-to-leading
order''~\cite{Campbell:2005bb}, namely the $b$ quark that appears
in the diagrams of fig.~\ref{fig:dr}, which accompanies the $b$ quark
coming from the decay of the top (not shown in fig.~\ref{fig:dr}).
In practice, this amounts to keeping events that satisfy 
\beq
\pt^b<\ptveto,
\label{ptveto}
\eeq
where appropriate. This means that in the case of partonic processes 
with no such additional $b$ quark, no veto is applied.
The reasoning for the condition in eq.~(\ref{ptveto}) is that 
harder $b$ quarks tend to have come from the decay of a top, and thus 
the probability for producing events with two hard $b$ quarks is
dominated by the contribution of diagrams with a $t$ and a $\bt$ both
on-shell or almost on-shell. By requiring the additional $b$ to be softer, one
thus reduces the interference between the doubly resonant diagrams and the
singly resonant diagrams which are identified with the $Wt$
channel. Furthermore in ref.~\cite{Campbell:2005bb}, the factorisation
scale is chosen to be equal to $\ptveto$. If this condition is not
met, doubly-resonant diagrams are removed from the computation
at the amplitude level.
Finally, processes with a $q\bq$ initial state are
not included, independently of the choice of factorisation scale.
This is only a problem of principle, since in practice the
numerical impact of such processes is so small that they can be neglected.
As in the other calculations discussed above, it is not possible
to apply this definition of the $Wt$ channel in an event 
generation/experimental situation, since the veto is based
on a partonic picture at a given order in $\as$.

\subsection{MC@NLO approach}
\label{sec:MCatNLOsol}
In this section, we consider the problem of defining the $Wt$ channel in
a way that is applicable in an event generator context, where both initial-
and final-state parton showers are present. Although none of the 
approaches discussed in subsection~\ref{sect:NLOcomp} can be
directly applied in this situation, they can be used to some degree 
in motivating suitable solutions.

The simplest and most drastic solution is to remove from the 
computation the contributions of those processes 
which contain doubly-resonant diagrams. At the NLO, this amounts
to removing {\em completely} diagrams having $gg$ or $q\bq$
initial states, i.e. regardless of whether or not they 
are doubly resonant. As far as the $q\bq$ contribution is concerned,
this is what is done in MCFM~\cite{Campbell:2005bb}. It is not quite
the same as what is done for the $gg$ initial state, but is numerically
very close to that, owing to the combination of the veto, and of
the choice of the factorisation scale $\muf=\ptveto$.
However, even if one generalises the definition of the veto
to apply in a parton shower context (as we will do in the following),
the removal of processes characterised by a particular initial
state is theoretically unfeasible. Firstly, renormalisation group
invariance is violated such that the underlying NLO calculation has 
the same formal scale dependence as a LO one (thus undermining one of the 
motivations for using an NLO generator at all). Secondly, and perhaps 
more seriously, partonic
processes mix as soon as higher-order effects are considered, as happens
when initial-state showers are present. Thus, we discount this 
method as a viable means of separating out the $Wt$ channel in a full 
Monte Carlo generator and do not present further results from it here.

Instead, we present two definitions of the $Wt$ channel, 
that are designed in such a way that, by comparing
them, one can directly assess the impact of the interference
with $t\bt$. Thus, if the results from the two definitions agree,
we can be confident of having isolated the $Wt$ channel. Such an
agreement, when it occurs, is the result both of the definitions themselves, 
and of final-state cuts which may or may not be applied. It is 
important to stress that the definitions that we shall give are
meaningful even without any subsequent cuts. This engenders a
greater degree of flexibility in the practical investigation of
whether the $Wt$ channel is well defined or not.

Our two definitions can be summarised as follows:
\begin{enumerate}
\item Diagram Removal (DR). Here one simply removes all diagrams in the 
NLO $Wt$ amplitudes that are doubly resonant (i.e. those diagrams shown 
in figure
\ref{fig:dr}). 
\item Diagram Subtraction (DS). In this approach, one modifies the NLO $Wt$
cross section by implementing a subtraction term designed to cancel 
{\em locally} the $t\bar{t}$ contribution.
\end{enumerate}
Note that DR is different from the removal of $gg$- and $q\bq$-initiated
processes discussed previously, since diagrams with these initial states 
are kept if they are not doubly resonant. Note also that the DS procedure
is similar to what has been proposed in ref.~\cite{Tait:1999cf}, but the
construction of its local subtraction term involves some technical 
complications, which we discuss later.

In order to discuss in detail the DR and DS definitions, we 
introduce some notation. Let us denote by 
${\cal A}_{\ab}$ the ${\cal O}(\gw\as)$ amplitude of the process:
\beq
\alpha(p_1)+\beta(p_2)\,\longrightarrow\,t(k_1)+W(k_2)+\delta(k)\,.
\label{TLproc}
\eeq
The precise identity of parton $\delta$ is not relevant for what follows, 
so we omit it from the notation. We have:
\beq
\ampI=\ampI^{(Wt)}+\ampI^{(t\bt)}\,,
\label{eq:ampdef}
\eeq
where the two terms on the r.h.s. are the respective contributions of the 
singly- and doubly-resonant diagrams to the na\"{\i}ve $Wt$
cross section. In the computation of the cross section at NLO, the
following quantity appears:
\beqn
\abs{\ampI}^2&=&\abs{\ampI^{(Wt)}}^2
+2\Re\left\{\ampI^{(Wt)}\ampI^{(t\bt)^\star}\right\}
+\abs{\ampI^{(t\bt)}}^2
\\*&\equiv&
\singlyI+\interfI+\doublyI\,.
\eeqn
We stress that the terms $\interfI$ and $\doublyI$ are non-zero only in
the cases of $\{\alpha,\beta\}=\{g,g\}$ and $\{q,\qb\}$, where
in the latter process $q$ may also be a $b$ quark. In these cases,
$\delta$ will be a $\bb$ quark or another down-type antiquark according 
to the CKM matrix. Furthermore,
$\doublyI$ has neither soft nor collinear singularities, while those of 
$\interfI$ are integrable and subleading w.r.t. those of $\singlyI$.
Therefore, in the context of an NLO computation in the FKS formalism 
(or for that matter in any subtraction formalism), the NLO real-emission 
contribution to the subtracted short-distance partonic cross section 
including the flux factor and phase space will be:
\beq
d\hat{\sigma}_{\ab}=\frac{1}{2s}
\left(\ssinglyI+\interfI+\doublyI\right)d\phi_3\,,
\label{dsig1}
\eeq
where the hat denotes that infrared singularities
have been suitably subtracted. It is understood that the subtraction is
performed by means of plus-type distributions, which therefore may apply
to the phase space as well. The hadroproduction 
cross section resulting from eq.~(\ref{dsig1}) is
\beqn
d\sigma&=&d\sigma^{(2)}+\sum_{\ab}\int dx_1 dx_2\lumiI d\hat{\sigma}_{\ab}
\nonumber\\*&=&
d\sigma^{(2)}+\sum_{\ab}\int \frac{dx_1 dx_2}{2x_1 x_2 S}\lumiI
\left(\ssinglyI+\interfI+\doublyI\right)d\phi_3\,,
\label{dsig2}
\eeqn
with $S$ the squared centre of mass energy of the colliding hadrons
and ${\cal L}_{\alpha\beta}$ the parton-level luminosity.
The quantity $d\sigma^{(2)}$, requiring a two-body phase space, 
denotes all contributions to the cross
sections that are not already included in eq.~(\ref{dsig1}), i.e.
the Born, soft-virtual, and collinear remainder terms.
Both $d\sigma^{(2)}$ and eq.~(\ref{dsig1}) (but not their sum,
eq.~(\ref{dsig2})) are convention-dependent, since finite pieces can 
be freely moved from one contribution to the other, but this is 
irrelevant for the following discussion. When the NLO computation
is then matched to parton showers according to the MC@NLO prescription,
the above equation must be modified by the subtraction of MC counterterms.
We can choose to absorb these in $\ssinglyI$, because this
is the only piece that contains leading soft and collinear singularities.
Thus the schematic form of eq.~(\ref{dsig2}) applies at both the NLO
and MC@NLO levels. In this notation, the DR cross section corresponds to:
\beq
d\sigma^{(\DR)}=d\sigma^{(2)}+
\sum_{\ab}\int \frac{dx_1 dx_2}{2x_1 x_2 S}
\lumiI\ssinglyI d\phi_3\,,
\label{dsigDR}
\eeq
i.e. there are now no terms $\interfI$ or $\doublyI$, as all doubly
resonant diagrams have been removed from the {\em amplitude}. As mentioned
previously, this cross section violates gauge invariance; this
issue will be discussed in sect.~\ref{sect:gaugedep}.

Starting from eq.~(\ref{dsig2}), we also define the DS cross section.
This amounts to writing:
\beqn
d\sigma^{(\DS)}=d\sigma-d\sigma^{subt}\,,
\label{eq:DS0}
\eeqn
where $d\sigma^{subt}$ is designed to remove numerically the doubly-resonant
contribution. This may be achieved locally by defining
\beqn
d\sigma^{subt}&=&\sum_{\ab}\int dx_1 dx_2\,
\lumiI\, d\sigma_{\ab}^{subt}\,;
\\*
d\sigma_{\ab}^{subt}&=&
\frac{1}{2s}\wdoublyI d\phi_3\,,
\label{ttsubt2}
\eeqn
such that the quantity
\beq
\doublyI-\wdoublyI
\label{Ddiff}
\eeq
will vanish when $M_{\bb W}^2\equiv (k+k_2)^2\to m_t^2$. 
Note that $\doublyI$ 
and $\wdoublyI$ themselves will, in such a limit, either diverge, if
$\Gamma_t=0$, or have a Breit-Wigner-like peak, if $\Gamma_t\ne 0$. 
The DS cross section in eq.~(\ref{eq:DS0}) can now be re-written 
in the same form as eq.~(\ref{dsigDR}):
\beq
d\sigma^{(\DS)}=d\sigma^{(2)}+
\sum_{\ab}\int \frac{dx_1 dx_2}{2x_1 x_2 S}
\lumiI\left(\ssinglyI+\interfI+\doublyI-\wdoublyI\right)d\phi_3\,.
\label{dsigDS}
\eeq
One sees that the difference between the DR and DS cross sections 
has the form:
\beq
d\sigma^{(\DS)}-d\sigma^{(\DR)}=
\sum_{\ab}\int \frac{dx_1 dx_2}{2x_1 x_2 S}
\lumiI\left(\interfI+\doublyI-\wdoublyI\right)d\phi_3\,,
\label{DSmDR}
\eeq 
and thus is composed of a contribution from the interference term, and of the
difference between the subtraction term and the true doubly resonant
contribution to the NLO cross section. 

Our aim is now to construct a gauge-invariant subtraction term, such 
that the difference $\doublyI-\wdoublyI$ is as close to zero as
possible. Note also that requiring the subtraction term to be
local and gauge invariant prevents this difference from being
identically zero.
The subtraction term should have the schematic form:
\begin{equation}
  \label{eq:17}
\wdoublyI = |{\cal A}^{(0)}(\ab\rightarrow t\bar{t})|^2 
\times \mathrm{BW}(M_{\bb W})
\times |{\cal A}^{(0)}(\bar{t}\rightarrow W\bar{b})|^2
\end{equation}
where ${\cal A}^{(0)}(\ab\rightarrow t\bar{t})$ is the LO 
(i.e., ${\cal O}(\as)$) top pair production
amplitude, ${\cal A}^{(0)}(\bar{t}\rightarrow W\bar{b})$ the decay amplitude of
the anti-top, and $\mathrm{BW}(M_{\bb W})$ is the
Breit-Wigner function.

The first obvious difficulty in constructing a workable implementation of 
eq.~(\ref{eq:17}) is that the kinematics
on the l.h.s. should be that of the full $\ab\rightarrow tW\bb$ process
(due to the requirement of local cancellation in eq.~(\ref{dsigDS})), 
whereas at the same time the $\bt$ needs to be on-shell in order to compute
the quantity ${\cal A}^{(0)}(\ab\rightarrow t\bt)$ in a gauge-invariant way.
A second difficulty is that, 
while ${\cal A}^{(0)}(\ab\rightarrow t\bt)$ is computed with top 
width $\Gamma_t=0$, the Breit-Wigner factor in eq.~(\ref{eq:17}) requires 
non-zero $\Gamma_t$.

As a first attempt to overcome these difficulties one can reshuffle the
momenta of the decay products, to obtain an on-shell $\bt$ quark. 
In this case, one uses a non-zero top width only
in the Breit-Wigner factor. However, a more fundamental problem with
eq.~(\ref{eq:17}) is that spin correlations are not included, thus spoiling
again the local cancellation property we seek to achieve.
Therefore, rather than eq.~(\ref{eq:17}), we need a subtraction term 
of the form:
\beq
  \label{eq:18}
\wdoublyI=\abs{\ampI^{(t\bt)}}^2_{\rm reshuffled}\,,
\eeq
where $\ampI^{(t\bt)}$ is defined in eq.~(\ref{eq:ampdef}).
In other words, $\wdoublyI$ would be identical to $\doublyI$, were
it not for the reshuffling. Being a full amplitude, eq.~(\ref{eq:18})
does implement spin correlations in the decay of the $\bar{t}$.
A divergence present in eq.~(\ref{eq:18}) when $\Gamma_t=0$ 
disappears by setting $\Gamma_t \neq 0$.
However, the reshuffling implies that $\wdoublyI$ is not likely
to have the Breit-Wigner shape in $M_{\bb W}$ that would be desirable
in order for the difference in eq.~(\ref{Ddiff}) to be as small as
possible in the whole phase space. This is easily rectified by 
defining:
\begin{equation}
  \label{eq:19}
\wdoublyI=\frac{BW(M_{\bb W})}{BW(m_t)}\,
\abs{\ampI^{(t\bt)}}^2_{\rm reshuffled}\,.
\end{equation}
By construction, the amplitude of this subtraction term at $M_{\bb W}=m_t$ 
is precisely such as to cancel the resonant contribution to
the NLO $Wt$ cross-section.

However, even after this modification, there is a problem. For appropriate
cancellations of collinear singularities, 
we would also have to use $\Gamma_t \neq 0$ in the
computation of $\abs{\ampI}^2$. 
Unfortunately, after setting $\Gamma_t \neq 0$ in the
radiative amplitudes in this way, their collinear limits are modified,
and in order not to disrupt the local cancellation of collinear divergences 
one would need to perform the full calculation adopting a framework for
the consistent inclusion of finite-width effects in all kinematic
regions of interest (e.g. the complex mass scheme~\cite{Denner:2005fg}).
Given the scope of the present paper, a more 
pragmatic approach suffices, which has been extensively used at LEP 
in dealing with resonant decays. Namely, we set $\Gamma_t \neq 0$ only 
in the doubly-resonant diagrams, and leave $\Gamma_t = 0$ in the 
singly-resonant ones. This strategy is straightforward to 
implement in \texttt{MADGRAPH}.

\section{Results}
\label{results}
In the previous section we discussed the problem of interference between 
$t\bt$-like amplitudes (followed by decay of the $\bt$) and 
$Wt$-like amplitudes. Two separation mechanisms, DR and DS, were defined
as being suitable for an all orders computation of the scattering amplitude,
as occurs in a Monte Carlo event generator. We have implemented both of these
separation mechanisms in the MC@NLO framework, as discussed in 
sect.~\ref{Wtprod}. We stress that the MC subtraction terms given
in sect.~\ref{Wtprod} are identical for DS and DR, since such terms 
only modify (w.r.t. a pure-NLO computation) the form of $\ssinglyI$, 
which is identical in the DS and DR cross sections.
In this section we present sample results, and compare in
detail the output of the DS and DR calculations. Our aim is to analyse,
from a perturbative point of view, the degree to which separation of 
the $Wt$ channel is possible, thus giving an upper bound for the
impact of the interference that can be obtained with a realistic
analysis. Detailed phenomenological results will not be presented here, 
but are postponed to a forthcoming publication. In particular, we only
consider fully leptonic decays of the final-state $W$ bosons, as these are
sufficient to furnish a comparison between the different separation
mechanisms. In practice, semileptonic rather than fully leptonic 
decays will be studied first by experiments at the LHC.
Furthermore, our results are obtained by neglecting production
spin correlations, which is not restrictive as far as comparing
the two definitions of the $Wt$ channel given here is concerned. 
We have however implemented spin correlations in the DR
calculation, using the method of ref.~\cite{Frixione:2007zp}. 
Their implementation in the DS calculation along those lines is
technically slightly more complicated (but possible), owing to the 
negativity of the ``squared'' matrix elements at some phase-space points 
(due to the subtraction term), and is also deferred to the future.

All of the following numerical results have been obtained for the LHC with the
MRST2002 default PDF set \cite{Martin:2002aw}, setting the top mass and
width to $m_t=170.9$ GeV and $\Gamma_t = 1.41$ GeV, as well as the $W$ mass and
width to $m_W=80.4$ GeV and $\Gamma_W = 2.141$ GeV. The default values
of the renormalisation and factorisation scales are equal to the top mass.
The LO results quoted in this section have been obtained using the
same parameters (including PDFs and two-loop $\as$) as those adopted
for NLO computations.

\subsection{Transverse momentum veto}
\label{sec:veto}
As discussed previously, the definitions we gave of the $Wt$ channel are
independent of any cuts that are subsequently used to further reduce
the interference from $t\bt$ production. Nevertheless, it is unrealistic
to define the $Wt$ channel, in an actual experimental environment, with
no cuts\footnote{We mean here cuts whose {\em only purpose} is that of 
separating the $Wt$ and $t\bt$ processes, and not generic selection cuts
aimed at selecting the $t$ and/or the $W$ in a collision event.}
at all. It is not the purpose of this paper to undertake a
thorough investigation of various cut strategies. Therefore, we shall
limit our considerations to one particular cut, motivated by the 
transverse momentum veto proposed in ref.~\cite{Campbell:2005bb},
and discussed here in sect.~\ref{sect:NLOcomp}. As stated there,
the veto acts on the additional $b$ quark that may be present at
the NLO level. Here we generalise this idea to an event generator
context i.e. to a situation in which we cannot tell with certainty
which parton or hadron is associated with the additional $b$ quark.

Firstly, one searches all $b$-flavoured hadrons in a given event,
and orders them in transverse momentum (or transverse energy).
All hadrons whose pseudorapidity $\eta_B$ is outside a given range,
which we choose to be
\beq
\abs{\eta_B}\le 2.5\,,
\label{eq:psrapveto}
\eeq
are ignored. Secondly, a veto analogous to that of eq.~(\ref{ptveto}) 
is applied on the transverse momentum (or transverse energy) of 
the second hardest $B$ hadron satisfying eq.~(\ref{eq:psrapveto}):
\beq
\pt^B<\ptveto.
\label{ptBveto}
\eeq
In cases where no second-hardest $B$ hadron satisfying the
above requirements can be found, the event is accepted.
Note that this is the case in processes which do interfere
 with $t\bt$ production, but in which a $b$
quark is replaced by another down-type quark -- an effect
off-diagonal in the CKM matrix.
In this paper, we assume a 100\% $b$-tagging efficiency.
However, it is clear that the veto procedure proposed here 
can be applied in more realistic $b$-tagging scenarios.
A more detailed phenomenological investigation of this issue is postponed
to a future publication. 

In a realistic analysis the veto would be accompanied
by a number of further cuts. For example, in the semileptonic decay
mode, the veto can be used in combination with a jet topology cut,
to further reduce the contamination of the $Wt$ ``signal'' from $t\bt$ 
``background''. Given that such cuts are absent in the results presented
here, our findings correspond to a somewhat pessimistic scenario
as far as the purity of the $Wt$ signal is concerned.

The definition of the veto adopted here can also be used in the
context of a pure-NLO, parton-level computation, such as that
of ref.~\cite{Campbell:2005bb}, by simply replacing $B$ hadrons
with $b$ quarks\footnote{$b$ quarks can also be considered in
an MC context. We have found that MC@NLO results obtained by imposing
$b$ quark vetos are very similar to those obtained with $B$ hadron vetos.}.
It is instructive to see, however, that the veto is a much more natural
constraint in an event generator context than in a parton-level,
fixed-order computation.
\begin{table}[htb]
\begin{center}
\begin{tabular}{|c|c|c|c|c||c|c|c|c|c|}\hline
\multicolumn{5}{|c||}{MC@NLO} &
\multicolumn{5}{c|}{NLO} \\\hline
$\sigma^{(\DR)}$ & $\sigma^{(\DS)}$ & $\sigma^{(\LO)}$ & 
$K^{(\DR)}$ & $K^{(\DS)}$ &
$\sigma^{(\DR)}$ & $\sigma^{(\DS)}$ & $\sigma^{(\LO)}$ & 
$K^{(\DR)}$ & $K^{(\DS)}$ \\\hline\hline
\multicolumn{10}{|c|}{$\ptveto=10$~GeV}\\\hline
34.66 & 33.89 & 26.60 & 1.30 & 1.27 & 
35.05 & 34.74 & 34.67 & 1.01 & 1.00 \\\hline\hline
\multicolumn{10}{|c|}{$\ptveto=30$~GeV}\\\hline
41.86 & 40.74 & 31.85 & 1.31 & 1.28 & 
39.93 & 39.67 & 34.67 & 1.15 & 1.14 \\\hline\hline
\multicolumn{10}{|c|}{$\ptveto=50$~GeV}\\\hline
44.61 & 42.92 & 33.71 & 1.32 & 1.27 & 
42.81 & 42.00 & 34.67 & 1.23 & 1.21 \\\hline\hline
\multicolumn{10}{|c|}{$\ptveto=70$~GeV}\\\hline
45.63 & 43.65 & 34.31 & 1.33 & 1.27 & 
44.41 & 42.90 & 34.67 & 1.28 & 1.24 \\\hline\hline
\multicolumn{10}{|c|}{$\ptveto=\infty$}\\\hline
46.33 & 44.12 & 34.67 & 1.34 & 1.27 & 
46.33 & 44.12 & 34.67 & 1.34 & 1.27 \\\hline
\end{tabular}
\end{center}
\caption{\label{tab:vetos} 
Results for the total DR, DS, and leading order cross sections (in pb), 
obtained with MC@NLO (five left columns) and our pure-NLO parton level 
computation (five right columns). See the text for details.
The notation $\ptveto=\infty$ denotes no veto at all.
}
\end{table}
In table~\ref{tab:vetos} we present the results for the DR and DS
total cross sections, obtained with MC@NLO and with our pure-NLO parton 
level computation. We also give the results (denoted by $\sigma^{(\LO)}$)
obtained in the two frameworks by keeping only LO matrix elements.
In the case of MC@NLO, $\sigma^{(\LO)}$ is thus equal to what 
one would get by simply running HERWIG standalone\footnote{In fact,
$Wt$ production is not implemented in HERWIG. We simply computed the
LO matrix elements, and used the Les Houches interface~\cite{Boos:2001cv} 
to give HERWIG the hard events, as in the case of MC@NLO.}. The table 
finally reports the values of the ratios of NLO over LO cross sections:
\beq
K^{(\DR)}=\frac{\sigma^{(\DR)}}{\sigma^{(\LO)}}\,,\;\;\;\;\;
K^{(\DS)}=\frac{\sigma^{(\DS)}}{\sigma^{(\LO)}}\,.
\eeq
In the fixed-order computation at LO, there is simply no second-hardest
$b$ quark, and $\sigma^{(\LO)}$ is therefore independent of the value
of $\ptveto$. On the other hand, in an MC context the initial-state $b$
quark entering the hard partonic process at LO eventually results, because of 
parton showers, in the generation of a $B$ hadron. Therefore, $\sigma^{(\LO)}$
computed with an event generator does depend on $\ptveto$. The interesting 
thing about this dependence is that it appears to be the same as that
obtained at the NLO, as can be inferred from the basically 
constant values of $K^{(\DR)}$ and $K^{(\DS)}$ obtained with MC@NLO. 
This is not the case for the parton-level fixed-order computation, 
where the factors $K$ display a significant dependence upon $\ptveto$. 
This raises the following two issues. At small-$\pt$, the fixed-order results
lack Sudakov suppression at $\pt^b\to 0$; it follows that the predictions 
obtained for small values of $\ptveto$ are not particularly meaningful.
At larger $\pt$'s, the real-emission matrix elements dominate, and an
NLO computation is expected to be reliable. Unfortunately, since the veto 
does not enter the LO results, it is not possible to sensibly estimate 
the impact of higher-order corrections: small changes to $\ptveto$ lead
to large changes in the factors $K$, which may effectively mask the
impact of $t\bt$ interference.
In summary, the imposition of a veto appears to be somewhat problematic
in parton-level, fixed-order computations. This issue would imply a much 
(perhaps overly) larger theoretical systematic error, relative
to an MC-based simulation, in the comparison between 
predictions and data.

\subsection{Gauge (in)dependence of DR cross sections}
\label{sect:gaugedep}
As pointed out in sect.~\ref{sec:MCatNLOsol}, the definition of the
DR cross section, in which certain diagrams are removed from 
an amplitude, is not gauge invariant. In this section, we will
argue that this is not a problem in practice, by repeating the DR
calculation in a number of alternative gauges. We have checked 
several different observables, namely the total cross section;
single-inclusive rapidity and transverse momentum distributions
of the $t$, $W$ and their decay products; and azimuthal and transverse
momentum correlations between the charged leptons. We present here
results for the total cross section and the transverse momentum $\ptll$
of the charged lepton pair. These are representative of the corresponding
results for other observables.

Our original calculation of the NLO cross section in the DR approach was
carried out in the Feynman gauge, where the gluon propagator is given by:
\begin{equation}
D_{\mu\nu}=-\frac{g_{\mu\nu}}{k^2}\,.
\label{prop1}
\end{equation}
The sum over the polarisation degrees of freedom of initial-state gluons
has been restricted to transverse polarisations.
In order to check the sensitivity to the choice of gauge, we repeated the
calculation in various covariant and non-covariant gauges. We summarise 
these results as follows.

\begin{figure}
\begin{center}
\scalebox{0.7}{\includegraphics{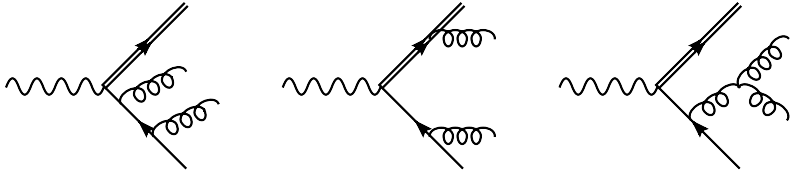}}
\caption{Diagrams kept in the $gg$ channel after removing resonant
contributions, shown before crossing. These formally represent also
the two additional diagrams in which the two gluons arising
from the fermion line have their momenta exchanged.}
\label{diags}
\end{center}
\end{figure}

First, we note that the gauge dependence can only enter in the $gg$
channel. This is because only diagrams involving two external gluons have
broken gauge invariance, and these occur in the $qg$ and $gg$
channels. However, the $qg$ channel has no resonant diagrams, thus no diagrams
are removed. The diagrams kept in the $gg$ channel (before crossing) are shown
in figure \ref{diags}.
Only the third diagram in figure \ref{diags} leads to a gauge dependence, due
to the presence of the gluon propagator. One can consider the general family
of covariant gauges that result from replacing the propagator of 
eq.~(\ref{prop1}) with:
\begin{equation}
D_{\mu\nu}=-\frac{1}{k^2}\left(g_{\mu\nu}+(1-\lambda)\frac{k_\mu
k_\nu}{k^2}\right).
\label{prop2}
\end{equation}
This is useful, as $\lambda=1$ reproduces the Feynman gauge calculation. One
thus has a continuous parameter in the DR calculation, such that it tends
smoothly to the Feynman gauge calculation in a well-defined limit. However, it
is easily seen from the form of the three gluon vertex that the second term in
eq.~(\ref{prop2}) decouples from the amplitude. Thus, DR results do
not depend on the choice of gauge within the family of covariant gauges.

One may also consider non-covariant gauges, in which the gluon propagator 
is given by\footnote{The gluon propagator also depends in general on 
a gauge-fixing parameter $\alpha$, which we set equal to zero in what 
follows. See e.g. refs.~\cite{Leibbrandt:1987qv,Bassetto:1991ue}.}:
\begin{equation}
D_{\mu\nu}=\left[-g_{\mu\nu}+\frac{n_\mu k_\nu+n_\nu k_\mu}{n\cdot
k}-\frac{n^2k_\mu k_\nu}{(n\cdot k)^2}\right]\frac{1}{k^2}.
\label{prop3}
\end{equation}
The third term decouples for the same reason as in a covariant gauge, but a
gauge dependence results from the second term. Here we report the numerical
results for the MC@NLO cross section in DR, obtained with non-covariant
gauges, for the three choices $n^2>0$, $n^2=0$, and $n^2<0$. We start in
table~\ref{tab:gauge} with the results obtained by integrating over the whole
phase space, except perhaps for the veto imposed on the second-hardest 
$B$ hadron of the event (see sect.~\ref{sec:veto}).
For reference, we also report the results obtained in the Feynman
gauge (which is our default).  We define a relative difference as follows:
\beq
\delta=10^{3}\times \frac{\sigma^{(\DR)}({\rm non-covariant})-
\sigma^{(\DR)}({\rm covariant})}{\sigma^{(\DR)}({\rm covariant})}\,.
\label{delta0}
\eeq
\begin{table}[htb]
\begin{center}
\begin{tabular}{|c|c|c|c|c|c|c|c|c|c|}\hline
\multicolumn{2}{|c|}{$\ptveto=10$} &
\multicolumn{2}{|c|}{$\ptveto=30$} &
\multicolumn{2}{|c|}{$\ptveto=50$} &
\multicolumn{2}{|c|}{$\ptveto=70$} &
\multicolumn{2}{|c|}{$\ptveto=\infty$}\\\hline
$\sigma$ & $\delta$ & $\sigma$ & $\delta$ & 
$\sigma$ & $\delta$ & $\sigma$ & $\delta$ &
$\sigma$ & $\delta$ \\\hline\hline
\multicolumn{10}{|c|}{Covariant}\\\hline
$34.66$ & $0$ &
$41.86$ & $0$ &
$44.61$ & $0$ &
$45.63$ & $0$ &
$46.33$ & $0$ \\\hline\hline
\multicolumn{10}{|c|}{$n^2>0$}\\\hline
$34.70$ & $1.15$ &
$41.90$ & $0.96$ &
$44.70$ & $2.01$ &
$45.73$ & $2.19$ &
$46.42$ & $1.94$ \\\hline\hline
\multicolumn{10}{|c|}{$n^2=0$}\\\hline
$34.75$ & $2.60$ &
$41.94$ & $1.91$ &
$44.69$ & $1.79$ &
$45.71$ & $1.75$ &
$46.42$ & $1.94$ \\\hline\hline
\multicolumn{10}{|c|}{$n^2<0$}\\\hline
$34.70$ & $1.15$ &
$41.87$ & $0.24$ &
$44.61$ & $0$ &
$45.63$ & $0$ &
$46.37$ & $0.86$ \\\hline
\end{tabular}
\end{center}
\caption{\label{tab:gauge} 
Gauge dependence of DR rates. The cross sections $\sigma$ are given in pb,
and veto values $\ptveto$ in GeV. Note the factor $10^3$ in the definition 
of $\delta$, eq.~(\ref{delta0}). The notation $\ptveto=\infty$ denotes 
no veto at all.
}
\end{table}
We observe very small relative differences, which are actually compatible 
with zero within the statistical errors of our runs (each of which 
consists of 500k events). Note that this conclusion holds regardless
of whether a veto is imposed or not.

\begin{figure}
  \begin{center}
      \epsfig{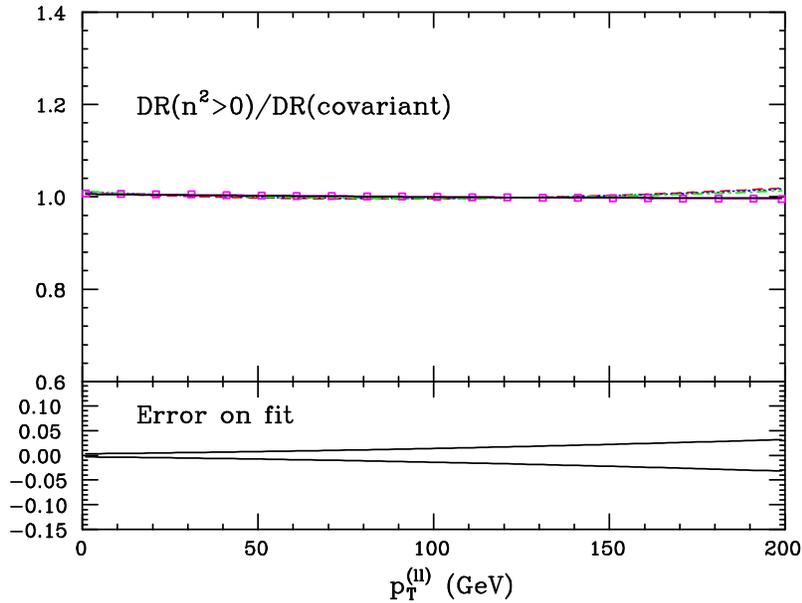}
\caption{\label{fig:axl1d_o_dr1} 
Gauge dependence of the DR cross section, as a function of the $\pt$
of the lepton pair, for a non-covariant gauge with $n^2>0$. Upper
pane: results of the fits, using the form in eq.~(\ref{fitform}).
Lower pane: envelope of curves obtained by varying all parameters
of the fit within their error ranges (see the text for details),
divided by the best fit curve, minus one. Black solid, red dashed, 
blue dotted, and green dot-dashed lines correspond to $\ptveto=10$, 
30, 50, and 70 GeV respectively. The magenta solid line with
open boxes is obtained without imposing any veto.
}
  \end{center}
\end{figure}

We have also checked that the same conclusion applies to more
exclusive observables. It is obviously impossible to reach the same
level of statistical accuracy as that in table~\ref{tab:gauge} for a 
differential distribution, especially in the tails of the transverse momentum
spectra which fall rather steeply. We have computed the ratios
\beq
R({\cal O})=\frac{d\sigma^{(\DR)}}{d{\cal O}}({\rm non-covariant})\Bigg/
\frac{d\sigma^{(\DR)}}{d{\cal O}}({\rm covariant})
\label{rat}
\eeq
bin-by-bin for the observables ${\cal O}$ (we have considered rapidities
and transverse momenta, both for single-inclusive observables and for 
correlations), and found all ratios to be compatible with one. The
typical value of $\abs{R({\cal O})-1}$ is actually always much smaller
than the statistical error affecting this quantity. The latter
may be of the order of a few tens of percent in individual bins
in the tails of $\pt$ distributions. However, since bin-by-bin
fluctuations tend to integrate to zero if several contiguous bins
are considered, we also fitted ratios in eq.~(\ref{rat}) for 
those cases in which ${\cal O}$ is a transverse momentum, by
assigning arbitrarily the same 0.1\% relative errors to all bins,
with the functional form
\beq
a_1+a_2\pt+a_3\pt^2\,.
\label{fitform}
\eeq
This ensures that the tail of the $\pt$ distributions is treated
in the fit on the same footing as the peak region, thus dealing
in an efficient way with the problem of bin-by-bin fluctuations, and 
allowing us to uncover hints of non-flatness in $R(\pt)$ as a
function of $\pt$. In this way, we have again found that all
$R(\pt)$ we have considered are compatible with one in the whole
$\pt$ range. As an example, we present here the results obtained
for the transverse momentum $\ptll$ of the lepton pair, for the four veto
choices, and for the 
gauge $n^2>0$. The ratio $R(\ptll)$ are presented in the upper pane of
fig.~\ref{fig:axl1d_o_dr1}. The result is also given for the
case in which no veto at all is applied.
As one can see from the figure, all curves resulting from the fit
are remarkably flat (they cannot actually be easily distinguished).
Some of them may be seen to depart from one at the largest $\ptll$ 
values, but this behaviour is not statistically significant. To show
this, we present in the lower pane of fig.~\ref{fig:axl1d_o_dr1} the
envelope of the curves obtained by considering all combinations
of values $a_i=a_i^0\pm\Delta(a_i)$, with $a_i^0$ and $\Delta(a_i)$
being the best values and their fitting errors respectively (as given 
by MINUIT~\cite{MINUIT}). The envelope is then divided by the best fit
curve and unity is subtracted, so as to give an upper bound for 
the error affecting the fitting procedure. Only the result relevant to
$\ptveto=10$~GeV is presented in the lower pane of fig.~\ref{fig:axl1d_o_dr1},
since all the others are essentially identical.

We therefore conclude that, regardless of the observable studied, the impact
of gauge dependence in the DR computation can be safely neglected in the
numerical studies that follow.

\subsection{Impact of interference}
In order to gauge how much of the difference between DS and DR is
due to the interference term alone, it is useful to define the quantity: 
\beq
d\sigma^{(\NI)}=d\sigma^{(2)}+ \sum_{I}\int \frac{dx_1 dx_2}{2x_1 x_2 S}
\lumiI\left(\ssinglyI+\doublyI-\wdoublyI\right)d\phi_3\,,
\label{dsigNI}
\eeq
where the label NI stands for {\em non-interference}. Clearly
\beq
d\sigma^{(\DS)}-d\sigma^{(\NI)}=
\sum_{I}\int \frac{dx_1 dx_2}{2x_1 x_2 S}
\lumiI\interfI d\phi_3\,,
\label{DSmNI}
\eeq
which is then a direct estimate of the interference contribution,
free of contaminations due to doubly-resonant contributions which
are present in eq.~(\ref{DSmDR}). Although eq.~(\ref{DSmNI}) is
not a physical quantity, it is useful to compute it because of
the expression for $d\sigma^{(\DS)}-d\sigma^{(\DR)}$ in eq.~(\ref{DSmDR}). 
As can be seen there, this quantity is the interference term eq.~(\ref{DSmNI}),
plus the difference between the doubly-resonant contribution and
the subtraction term. Thus, eq.~(\ref{DSmNI}) would be identical
to $d\sigma^{(\DS)}-d\sigma^{(\DR)}$ in the limiting case
$\doubly\equiv\wdoubly$ (which, we recall, is impossible to achieve if local 
cancellation and gauge invariance are simultaneously imposed).

Results for the total cross sections after implementation in MC@NLO are shown
in table \ref{tab:interf}, where we have defined the relative differences:
\beqn
\delta_1&=&\frac{\sigma^{(\DS)}-\sigma^{(\DR)}}{\sigma^{(\DR)}}\,,
\label{deltadef1}
\\
\delta_2&=&\frac{\sigma^{(\DS)}-\sigma^{(\NI)}}{\sigma^{(\DR)}}\,.
\label{deltadef2}
\eeqn
\begin{table}[htb]
\begin{center}
\begin{tabular}{|c|c|c|c|c||c|c|}\hline
$\sigma^{(\DR)}$ & $\sigma^{(\DS)}$ & $\sigma^{(\NI)}$ & 
$\sigma^{(\DS)}-\sigma^{(\DR)}$ & $\delta_1$ &
$\sigma^{(\DS)}-\sigma^{(\NI)}$ & $\delta_2$\\\hline\hline
\multicolumn{7}{|c|}{$\ptveto=10$~GeV}\\\hline
34.66 & 33.89 & 35.08 & $-$0.77 & $-$2.2\% & $-$1.19 & $-$3.4\% \\\hline\hline
\multicolumn{7}{|c|}{$\ptveto=30$~GeV}\\\hline
41.86 & 40.74 & 42.83 & $-$1.12 & $-$2.7\% & $-$2.09 & $-$5.0\% \\\hline\hline
\multicolumn{7}{|c|}{$\ptveto=50$~GeV}\\\hline
44.61 & 42.92 & 46.59 & $-$1.69 & $-$3.8\% & $-$3.67 & $-$8.2\% \\\hline\hline
\multicolumn{7}{|c|}{$\ptveto=70$~GeV}\\\hline
45.63 & 43.65 & 48.24 & $-$1.98 & $-$4.3\% & $-$4.59 & $-$10.0\% \\\hline\hline
\multicolumn{7}{|c|}{$\ptveto=\infty$}\\\hline
46.33 & 44.12 & 49.58 & $-$2.21 & $-$4.8\% & $-$5.46 & $-$11.8\% \\\hline
\end{tabular}
\end{center}
\caption{\label{tab:interf} 
Results for the total DR, DS, and NI cross sections (in pb) defined in 
eqs.~(\ref{dsigDR}), (\ref{dsigDS}), and~(\ref{dsigNI}), for various values 
of the veto. We also give the relative differences as defined in 
eqs.~(\ref{deltadef1}) and~(\ref{deltadef2}). The notation 
$\ptveto=\infty$ denotes no veto at all.
}
\end{table}
Overall, we observe that the relative differences are not large, and decrease
for tighter vetoes. This is to be expected given that by requiring a stricter
veto, one filters out more of the $t\bar{t}$ process. 
Note that the relative differences due to the interference term alone are
larger than those between DR and DS, implying that the terms $\interf$ and
$\doubly-\wdoubly$ cancel out to some extent. 

Although not reported in
this table, we have studied the scale dependence of these results, 
by varying the renormalisation and factorisation scales between $m_t/2$ 
and $2m_t$. The two scales are always set equal to a common value,
which is somewhat restrictive but sufficient for our present purposes.
Although the individual cross sections depend very mildly on scales
(the variations w.r.t. the central values being $~^{+2}_{-3}$\%,
$~^{+1}_{-2}$\%, and $~^{+3}_{-4}$\% for DR, DS, and NI respectively),
this is not the case for $\sigma^{(\DS)}-\sigma^{(\DR)}$ and
$\sigma^{(\DS)}-\sigma^{(\NI)}$, which have variations w.r.t. their
central values of about $\pm 20$\% and $\pm 30$\% respectively.
This behaviour is not surprising: the individual cross sections have
been defined with the specific purpose of describing the $Wt$ cross
section, which is known from the literature to have a mild scale
dependence. On the other hand, as explicitly shown in eqs.~(\ref{DSmDR})
and~(\ref{DSmNI}), the differences of cross sections considered 
in table~\ref{tab:interf} are dominated by LO $t\bt$ production, which 
has a significant scale dependence (see e.g. ref.~\cite{Cacciari:2008zb}
for a recent update).

We have also carried out similar studies on differential distributions. 
The results presented in table~\ref{tab:interf} are, not surprisingly,
dominated by the small transverse momentum regions, where the bulk of
the cross section lies. Analogously to what was done in eq.~(\ref{rat}),
we have computed the ratios
\beq
R^{(\DS)}({\cal O})=
\frac{d\sigma^{(\DS)}}{d{\cal O}}\Bigg/
\frac{d\sigma^{(\DR)}}{d{\cal O}}\,,
\;\;\;\;\;\;\;\;
R^{(\NI)}({\cal O})=
\frac{d\sigma^{(\NI)}}{d{\cal O}}\Bigg/
\frac{d\sigma^{(\DR)}}{d{\cal O}}\,,
\label{ratDSNIDR}
\eeq
for several observables. In the cases in which ${\cal O}$ is a rapidity,
these ratios are flat over the whole kinematically-accessible range,
with values consistent with those given in table~\ref{tab:interf}.
The situation is more interesting if one considers transverse momenta,
since in such cases the ratios in eq.~(\ref{ratDSNIDR}) display a
non-trivial shape. In particular, the largest values of 
$\abs{R^{(\DS)}-1}$ and $\abs{R^{(\NI)}-1}$, amongst the
observables we have studied, are found in the tail of $\ptll$
(which is the reason why this observable has been considered as a 
case study in this paper). 

We start by presenting results for the quantities defined in
eq.~(\ref{ratDSNIDR}). In order to be able to superimpose five
curves on the same plot and keep visibility (since histograms 
would blur the picture because of bin-by-bin fluctuations in the
tail), the ratios have again been fitted with the functional form 
of eq.~(\ref{fitform}).
\begin{figure}[htb]
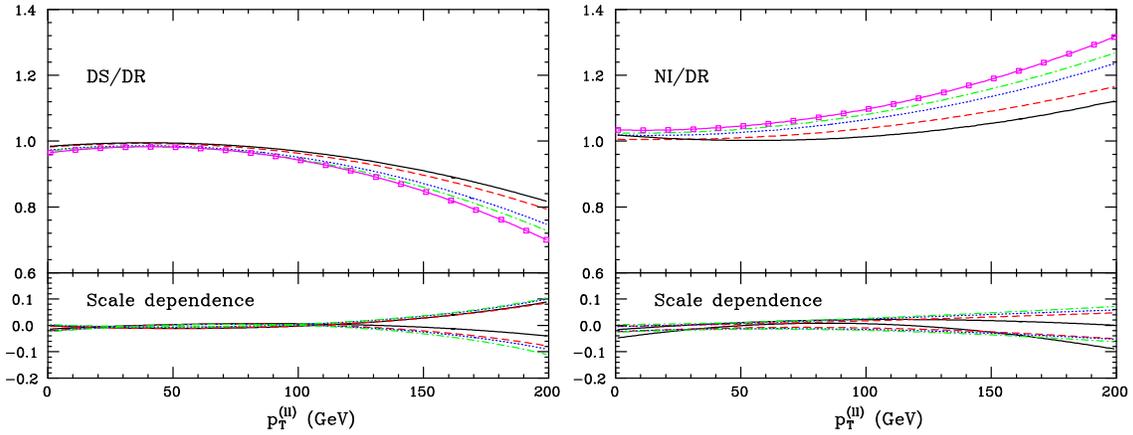

  \begin{center}
    \epsfig{figure=ptll2_ds1_o_dr1.eps,width=0.49\textwidth}
    \epsfig{figure=ptll2_ni1_o_dr1.eps,width=0.49\textwidth}
\caption{\label{fig:rDSrNI} 
Upper panes: results for the ratios defined in eq.~(\ref{ratDSNIDR}),
as a function of $\ptll$  and for various vetos. Lower panes: relative
scale dependence (see eq.~(\ref{Rscdep})). The linestyles are the same 
as those of fig.~\ref{fig:axl1d_o_dr1}.
}
  \end{center}
\end{figure}
The results are given in the upper panes of fig.~\ref{fig:rDSrNI}. 
As can be seen from the figure, the impact of interference can be
very large for large enough values of $\ptll$, regardless of the choice for
$\ptveto$ (although, of course, it is less significant for small vetos).
\begin{figure}[htb]
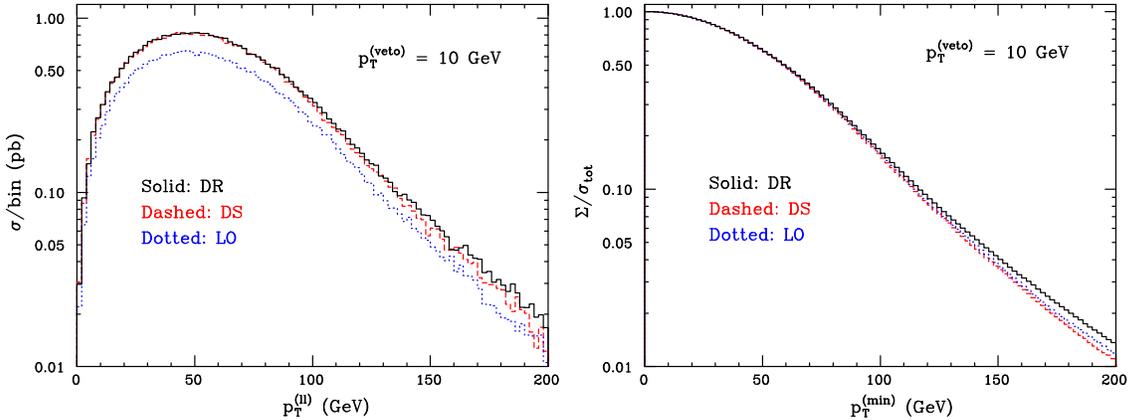

  \begin{center}
    \epsfig{figure=ptll_Bv10.eps,width=0.49\textwidth}
    \epsfig{figure=ptll_Bv10_int.eps,width=0.49\textwidth}
\caption{\label{fig:ptll} 
Left pane: differential DR, DS, and LO distributions in $\ptll$.
Right pane: integral of the same distributions in the range 
$\ptmin<\ptll<\infty$, divided by the respective total rates. These results 
are relevant to the case $\ptveto=10$~GeV.
}
  \end{center}
\end{figure}
It must be stressed however that, at large $\ptll$, the cross section is
small. This is documented in fig.~\ref{fig:ptll}, where we present in the
left pane the differential distributions in $\ptll$, as computed
with DR, DS, and at LO. The same information is presented in the
right pane of the figure, in an integral form:
\beq
\frac{1}{\sigma_{tot}}\Sigma(\ptmin)=\frac{1}{\sigma_{tot}}
\int_{\ptmin}^\infty d\ptll\frac{d\sigma}{d\ptll}\,.
\eeq
The results of fig.~\ref{fig:ptll} have been obtained by
choosing $\ptveto=10$~GeV. Although the absolute value of the
differential cross section has a non-negligible dependence
on $\ptveto$, its shape is relatively stable against variations 
of $\ptveto$. Thus, the right pane of fig.~\ref{fig:ptll} can
be used with table~\ref{tab:interf} for estimating the number of
events with $\ptll\ge\ptmin$ for any choice of the veto (or in
its absence).

The comparison between fig.~\ref{fig:rDSrNI} and fig.~\ref{fig:ptll} 
shows that, for the majority of events, the impact of interference
is moderate. On the other hand, these results imply that conclusions
concerning interference of $Wt$ with $t\bt$ are observable dependent.
One particular analysis may be sensitive to an observable which 
receives much larger contributions from $\interf$ and $\doubly-\wdoubly$ 
than the total rate. With fig.~\ref{fig:rDSrNI} 
we have presented a worst-case scenario among the observables studied.

It is also interesting to study the scale dependence of these results.
This cannot be done in the same way as for the total rates of
table~\ref{tab:interf}. There, differences of cross sections have
been considered, and the relative variation w.r.t. the central
values computed. In the case of the tail of a steeply falling
differential distribution, bin-by-bin differences lead to such small
numbers that it is impossible, in practice, to obtain a statistically
significant result for the relative variation, since the latter would
be a ratio of two extremely small numbers. On the other hand, we 
are able to compute the scale dependence of the ratios defined
in eq.~(\ref{ratDSNIDR}). In the lower panes of fig.~\ref{fig:rDSrNI},
we present the quantities
\beq
\frac{R^{(\DS)}\left(\ptll;\muf=\mur=\kappa m_t\right)}
{R^{(\DS)}\left(\ptll;\muf=\mur=m_t\right)}-1\,,
\;\;\;\;\;\;\kappa=1/2,2
\label{Rscdep}
\eeq
(and the analogous one for NI) as a function of $\ptll$. 
Equation~(\ref{Rscdep}) will not give directly the scale dependence 
of a $t\bt$-dominated cross section, as was the case for the 
differences of rates of table~\ref{tab:interf}, but rather that
of the fraction of the DS and NI cross sections
arising from $t\bt$-like contributions. As can be seen from the figure,
the scale dependence of such a fraction is largest in the region
where the interference between singly- and doubly-resonant
diagrams is largest. This is consistent with the na\"{\i}ve
expectation that $Wt$-like contributions have a much milder
scale dependence than $t\bt$-like ones. On the other hand, 
the study of scale dependence
shows that the qualitative conclusion that one draws from 
fig.~\ref{fig:rDSrNI} is perturbatively stable, and that the interference 
is significant only in the large-$\pt$ region.

\section{Discussion}
\label{discussion}
In this paper we have presented the first calculation of single-top
production in association with a $W$ boson, accurate to NLO in QCD
and interfaced with parton showers according to the MC@NLO formalism.
In studying the $Wt$ channel, our main aim has been to investigate 
whether or not one can define it as a separate production process, 
since only in the case of a positive answer is the computation of
the full NLO corrections meaningful and feasible with 
present technologies. The NLO+parton shower framework provides
an excellent platform for addressing the problem. Parton showers give
the capability of studying final states in a realistic environment, 
and testing ideas that can be applied without any modifications to
an experimental analysis. The underlying NLO computation is,
on the other hand, directly sensitive to interference effects
with $t\bt$ production.

We have considered two separation mechanisms for defining the $Wt$
channel: diagram removal (DR) and diagram subtraction (DS). We have
defined them in such a way that it is not necessary, in order to
generate full hadronic events with high efficiency, to apply any
kinematic cuts. In this way complete flexibility is achieved, to 
study the effects of any cuts whose aim is that of enhancing the
$Wt$ ``signal'' w.r.t the $t\bt$ ``background''. As an example,
we have considered only one such cut, namely a veto
on the transverse momentum of the second-hardest $B$ hadron
of the event.

The DR cross section is obtained by removing the doubly-resonant 
diagrams from the calculation of the $Wt$ {\it amplitude}, and thus violates 
gauge invariance. We have demonstrated that this is not a problem in 
practice. The DS cross section involves a modification of the na\"{\i}ve 
$Wt$ {\it cross-section} by a gauge invariant subtraction term which acts to 
remove the doubly-resonant contribution from $t\bt$ final states. 
Given that this is done at the cross-section level, an interference 
term between the $t\bar{t}$ and $Wt$ processes remains. 

Ultimately, we are in a position to decide whether it is possible,
and useful for data analysis, to define the $Wt$ channel as a production 
process in its own right. 
The key question is whether an approximate treatment of interference within
a higher-order computation has to be preferred, for the sake of a counting
experiment, to an exact treatment of interference without QCD corrections.
Furthermore, the separation of $Wt$ and $t\bt$ processes allows one
to consider NLO $Wt$ results (which are ${\cal O}(\gw^2\as^2)$) alongside
NLO $t\bt$ results (which are ${\cal O}(\as^3)$). This makes 
a realistic description of phenomenology possible (since
it is well known that higher-order corrections for $t\bt$ production
are crucial in this respect), in spite of being unable to treat
the problem exactly in its full complexity.
In this paper we have found that NLO corrections
are larger than interference effects, especially if cuts are employed
that are designed to enhance the $Wt$ signature. The answer to
the above question therefore appears to be {\em yes}. However, the
definitions that render it possible to compute higher-order corrections
also imply an ambiguity in the theoretical predictions, that we identify
with the difference between DR and DS cross sections. We have found
that such an ambiguity is observable-dependent, and in particular
is larger in those regions of the phase space that correspond to 
large transverse momenta. Since this ambiguity will be directly
related to the theoretical systematic error, it follows that the accuracy
with which $Wt$ properties can be measured depends on the observables
that are most relevant to a given analysis. 

To conclude, having implemented the $Wt$ production mode in an NLO plus parton
shower context, we find based on our subsequent analysis that it does indeed
seem feasible to analyse this process at the LHC.

\acknowledgments 
We would like to thank Wim Beenakker, John Campbell, Borut Kersevan, 
Fabio Maltoni, Jan Smit, Francesco Tramontano  and Giulia Zanderighi for 
valuable discussions. We would also like to thank the CERN TH division
for hospitality during the completion of this work. C.W. and E.L would
like to thank the Galileo Galilei Institute for Theoretical Physics for
the hospitality and the INFN for partial support during the completion
of this work. S.F. would like to thank Nikhef for hospitality and
support on many different occasions. The work of C.W.,
E.L. and P.M.  is supported by the Netherlands Foundation for
Fundamental Research of Matter (FOM) and the National Organization for
Scientific Research (NWO); that of B.W. is supported in part by the UK
Science and Technology Facilities Council. 
\appendix
\section{$W^- t$ and $W^+ \bt$ production}
\label{sec:wbart-prod-same}
In this appendix we demonstrate that the squared amplitudes for single
top production with an associated $W$ boson are independent of whether
the final state top is a quark or an antiquark.

Firstly there is the issue of a possible charge asymmetry due to the
electroweak coupling of the $W$ boson. This merely constrains the
helicity of the fermion line which includes the top quark in the
graph, which occurs either as an open line or a closed loop. The
$t\rightarrow\bar{t}$ transformation changes the direction of the
line, and leads to a constraint on the helicity of the corresponding
antifermion line. However, the sign and size of the coupling is the
same in both cases.

Furthermore, there is the issue of a possible QCD asymmetry. One must
consider in more detail the open or closed fermion loop associated
with the top quark (and the $b$ quark from top decay). The replacement
$t\rightarrow\bar{t}$ can affect the amplitude in two ways:
\begin{enumerate}
\item It changes the sign of all momenta which occur in uncut
  propagators along the top quark line\footnote{Cut propagators
    correspond to terms in the fermion trace of form $\slsh{p}\pm m_t$
    depending on whether quark or antiquark spinors are involved i.e.
    the sign of the momentum is not changed.}. If $p_i$ are the
  propagating momenta, the fermion trace in each diagram will contain
  the following terms:
\begin{equation}
(\slsh{p}_1-m_t)(\slsh{p}_2-m_t)\ldots(\slsh{p}_M-m_t),
\label{momenta}
\end{equation}
where $M$ is the number of uncut fermion momenta, and there may be
additional Dirac matrices between the propagator factors.
Interchanging top and antitop quarks gives instead the terms
\begin{equation}
(-\slsh{p}_1-m_t)(-\slsh{p}_2-m_t)\ldots(-\slsh{p}_M-m_t)=
(-1)^M(\slsh{p}_1+m_t)(\slsh{p}_2+m_t)\ldots(\slsh{p}_M+m_t).
\label{momenta2}
\end{equation}
Given that the number of Dirac matrices in each interference graph is
even, only terms involving an even number of masses survive. Then the
amplitude for antiquark production is related to that for quark
production via
\begin{align}
  {\cal A}_{\bar{t}}&=C_{diag}{\cal A}_t\notag\\
  &=(-1)^M{\cal A}_t,
\label{abar}
\end{align}
where ${\cal A}_{t,\,\bar{t}}$ is the amplitude associated with a
given interference diagram. Here $C_{diag}$ is the parity under
$t\rightarrow\bar{t}$ of the diagram i.e. $C_{diag}=\pm 1$.
\item The colour factor may be affected. In each of the diagrams for
  $Wt$ production, one always has
\begin{equation}
{\cal C}_{\bar{t}}=C_{col}{\cal C}_t,
\end{equation}
where ${\cal C}_{t,\,\bar{t}}$ are the colour factors for the
amplitude with a top and antitop quark respectively, and $C_{col}=\pm
1$.
\end{enumerate}
The total parity under $t\rightarrow\bar{t}$ is then given by
\begin{equation}
C=C_{diag}C_{col}=\pm 1.
\label{parity}
\end{equation}
Squared real emission diagrams will always have an even number of
uncut propagators, and also symmetric colour factors under
$t\rightarrow\bar{t}$. Regarding interference diagrams, there are 48
in total. These can be subdivided into $gg$, $bb$, $bg$ and $b\bar{b}$
initial states. Then diagrams with other quarks in the initial state
form a subset of those already specified. The $gg$ and $b\bar{b}$
diagrams are always associated with real emissions. For
the $bb$ (and, hence, $qq$) initial state, there is only one Feynman
amplitude and hence no interference term is possible. The $qg$ initial
states can be associated with real or virtual emissions. By evaluating 
the number of uncut
fermion propagators and the colour factor for each graph, one can find
its parity under $t\rightarrow\bar{t}$ using eq.~(\ref{parity}).

There are two types of diagrams:
\begin{enumerate}
\item Diagrams with no triple gluon coupling. These all have symmetric
  colour factors under $t\rightarrow\bar{t}$, and an even number of
  uncut fermion propagators. Hence $C=1$ for such graphs.
\item Diagrams with a triple gluon coupling. These all have
  antisymmetric colour factors, and an odd number of uncut fermion
  propagators (this latter fact can be easily appreciated by
  considering removing a gluon line from a fermion in graphs of type 1
  and reattaching it to a gluon line). Hence $C=(-1)^2=+1$ for these
  graphs.
\end{enumerate}
One finds that every graph is even upon replacing top quarks by antitop
quarks, and so the total squared amplitude for $Wt$ production is the same for
both $t$ and $\bar{t}$. 

Even if $W^- t$ and $W^+ \bt$ matrix elements are identical at this order,
there is still the possibility of a charge asymmetry arising from
differences between the $b$ and $\bar{b}$ parton densities\footnote{As
already discussed in refs.~\cite{Tait:1999cf,Campbell:2005bb}.}. However, 
these densities are equal in the global fits available at present.
Thus, overall, the cross-sections and
differential distributions considered here for top and anti-top
production are strictly equal.

\section{Calculation of $\tilde{\cal{M}}$ in the helicity formalism}
\label{sec:calc-tild-helic}
NLO subtraction formalisms such as FKS achieve a good numerical
stability through the local cancellation between large contributions of
opposite sign, which correspond to
the real-emission matrix elements in the soft and collinear regions,
and the counterterms. The counterterms are in fact constructed by
calculating the above limits of the matrix elements, whose form
is universal. If one considers e.g. the initial-state collinear 
limit $p_2\parallel k$ of the process
\beq
\alpha(p_1)+\beta(p_2)\,\longrightarrow\,X(K)+\delta(k)\,,
\label{abXc}
\eeq
with $X$ a set of final state particles which are not relevant here
(in our case, $X\equiv t+W$), the matrix element squared is
(see e.g. eq.~(B.41) of ref.~\cite{Frixione:1995ms})
\beq
\mat(p_1,p_2)\,\stackrel{p_2\parallel k}{\longrightarrow}\,
\frac{4\pi\as}{k\mydot p_2}\left[P(z)\mat^{(0)}(p_1,zp_2)+
Q(z)\tilde{\mat}(p_1,zp_2)\right]\,,
\label{Mclimit}
\eeq
where $P$ are the usual Altarelli-Parisi kernels, $Q$ are other universal
kernels (given at the leading order in eqs.~(B.42)--(B.45) of 
ref.~\cite{Frixione:1995ms} for initial-state collinear splittings, 
and in eqs.~(B.31)--(B.34) of that paper for final-state collinear splittings; 
these kernels are thus different for spacelike and timelike branchings already 
at the leading order), $\mat^{(0)}$ is the relevant Born contribution,
and $\tilde{\mat}$ is a Born-like function, which however keeps track
of the azimuthal correlations in the branching process. The contribution
of $Q\tilde{\mat}$ vanishes upon integration over the azimuthal angle
of the branching,
which is why this term can be neglected in the analytical computation of 
the collinear divergences in $4-2\epsilon$ dimensions. Locally, it is 
different from zero if the parton involved in the branching which enters
the hard reaction is a gluon, and therefore
needs to be taken into account for the construction of the local counterterms
in a (efficient) numerical NLO program.

\begin{figure}
\begin{center}
\scalebox{0.8}{\includegraphics{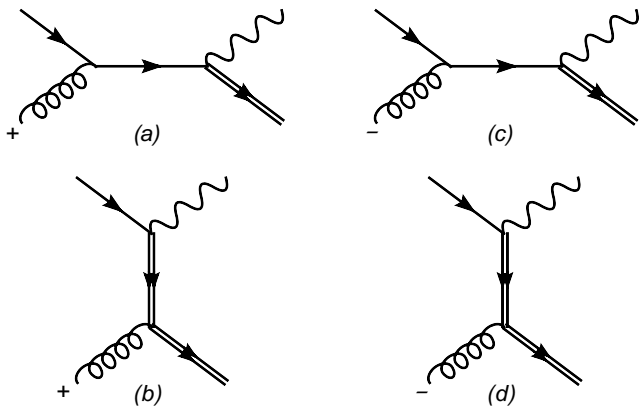}}
\caption{Diagrams used in the calculation of $\tilde{\cal M}$
for $0\rightarrow b(p_1)+g(p_2)+ W^-(p_3) + t(p_4)$, i.e. all momenta are
defined to be outgoing.}
\label{diagrams}
\end{center}
\end{figure}

Unlike the case of $s$- and $t$-channel single-top production,
the $Wt$ channel has a gluon entering the hard reaction at the Born level,
and we thus need to compute the relevant $\tilde{\mat}$ (which is
process dependent, whereas the kernels $Q$ are universal).
This term is (see eq.~(B.23) of ref.~\cite{Frixione:1995ms})
\begin{equation}
\tilde{\cal M}={\cal F}\real\left\{\frac{\langle k p_2\rangle}{[k p_2]} 
\ap^{(0)\dag}\am^{(0)}\right\},
\label{mt1}
\end{equation}
with ${\cal A}_\pm^{(0)}$ the Born amplitude for $Wt$ production with a
positive/negative helicity initial state gluon respectively, and ${\cal F}$ 
a term which includes the flux and spin- and colour-average factors,
and the coupling constants. We have computed
the r.h.s. of eq.~(\ref{mt1}) in two independent ways, and found
agreement. In this appendix we present some details on how the
computation can be carried out by using spinor helicity 
methods~\cite{Berends:1981rb,De Causmaecker:1981bg,Xu:1986xb,
Kleiss:1985yh,Gunion:1985vca}. As is conventional when applying such methods,
we take all momenta to be {\it outgoing} in this appendix, in contrast to the 
rest of the paper. Furthermore, we relabel the outgoing momenta $k_2$, $k_1$ 
as $p_3$ and $p_4$ respectively (see fig.~\ref{diagrams}), in order to be 
able to use a compact notation in what follows.

The relevant diagrams are illustrated in fig.~\ref{diagrams}. 
Besides the null momenta $p_1$ and $p_2$, we define the null vector
\begin{equation}
p_5=p_4-\frac{m_t^2}{u_1}p_2,
\label{p5def}
\end{equation}
where $u_1=u-m_t^2=(p_2+p_4)^2-m_t^2$. As is common, we adopt the notation
\begin{equation}
|k^\pm\rangle \equiv u_\pm(k),\quad \langle k^\pm|\equiv\bar{u}_\pm(k)
\label{spindefs}
\end{equation}
for massless quark spinors. Defining the Mandelstam invariants
\begin{equation}
s=(p_1+p_2)^2;\quad t=(p_1+p_4)^2\label{invars},
\end{equation}
the contributions $\ap^{(s)}$ and $\ap^{(u)}$ of the positive helicity 
diagrams (denoted by (a) and (b) respectively in
fig.~\ref{diagrams}) are given by (neglecting colour factors)
\begin{align}
\ap^{(s)}&=-\frac{2i\sqrt{2}}{s\langle 12\rangle
}\epsilon_\mu(p_3)\bar{u}(p_4)\gamma^\mu\ket{2-}\langle 21\rangle
\bra{2+}u(p_1);\label{aps}
\\* 
\ap^{(u)}&=\frac{2i\sqrt{2}}{u_1\langle 12\rangle}
\epsilon_\mu(p_3)\bar{u}(p_4)\left[\ket{1+}[25]\bra{5-}+m\ket{2-}\bra{1-}
\right]\gamma^\mu
u(p_1)\label{apt},
\end{align}
where $\epsilon_\mu(p_3)$ is the polarisation vector of the $W$ boson, and we
have used the standard notation
\begin{equation}
[ij]=\langle i+|j-\rangle ,\quad \langle ij\rangle =\langle i-|j+\rangle .
\label{defs}
\end{equation}
In deriving eqs.~(\ref{aps}) and~(\ref{apt}) we have used the following
expressions for the gluon polarisation vectors
\begin{equation}
\epsilon_{+,\mu}(p_2,q)=\frac{\langle q-|\gamma_\mu|2-\rangle}{\sqrt{2}\langle
q2\rangle},\quad \epsilon_{-,\mu}(p_2,q)=\frac{\langle
q+|\gamma_\mu|2+\rangle}{\sqrt{2}[2q]}.
\label{polars}
\end{equation}
Here $q$ is an arbitrary null reference momentum, which has been set
to $p_1$. One is in principle able to choose a different reference
momentum for the negative helicity diagrams, owing to the fact that
helicity amplitudes are separately gauge invariant. However, if one
also chooses $q_2=p_1$ in the negative helicity case then diagram (c)
in fig.~\ref{diagrams} vanishes. This relies on the helicity of the
light quark line being fixed by the chiral boson coupling. Then the
negative helicity amplitude is given solely by diagram (d):
\begin{equation}
\am^{(0)}=
\frac{2i\sqrt{2}}{u_1[21]}\epsilon_\mu(p_3)\bar{u}(p_4)\left[\frac{u}{u_1}
\ket{2+}[12]\bra{2-}+\ket{2+}[15]\bra{5-}+m_t\ket{1-}\bra{2-}\right]\gamma^\mu
u(p_1).
\label{am}
\end{equation}
It can be checked that $|\ap^{(0)}|^2+|\am^{(0)}|^2$, 
after summing over the $W$ boson and
quark spins, is equal to the known Born result, eq.~(\ref{eq:3}). 
For the helicity interference term, one uses eqs.~(\ref{aps}), (\ref{apt}),
and~(\ref{am}) to find ($\ap^{(0)}=\ap^{(s)}+\ap^{(u)}$):
\begin{equation}
\ap^{(0)\dag}\am^{(0)}=\frac{8}{u_1^2m_W^2}\frac{[15]^2\langle 25\rangle
^2}{[12]^2}(m_W^2-m_t^2)(2m_W^2+m_t^2).
\label{apam}
\end{equation}
Note the presence of squared spinor products, which cannot immediately be
evaluated to form dot products. This is to be expected, given that
$\tilde{\cal M}$ is not a Lorentz invariant quantity. To evaluate the
quantities in eq.~(\ref{apam}) one can parameterise 4-momenta in the center of
mass frame of the incoming particles as follows:
\begin{align}
p_1&=\left(\frac{\sqrt{s}}{2},0,0,\frac{\sqrt{s}}{2}\right);\label{p1}\\
p_2&=\left(\frac{\sqrt{s}}{2},0,0,-\frac{\sqrt{s}}{2}\right);\label{p2}\\
p_5&=\left(E,E\sin\theta\cos\phi,E\sin\theta\sin\phi,E\cos\theta\right).
\label{p5}
\end{align}
Basis spinors satisfying the massless Dirac
equation with conventional normalisation $u^\dag(p_i)u(p_i)=2p_i^0$
are (choosing arbitrary phases)
\begin{equation}
u_-(p_5)=\left(\begin{array}{c}0\\0\\-\sqrt{E(1-\cos\theta)}e^{-i\phi}\\
\sqrt{E(1+\cos\theta)}\end{array}\right);
\quad u_\lambda(p_1)=\left(\begin{array}{c}s^{\frac{1}{4}}\delta_{\lambda
+}\\0\\0\\s^{\frac{1}{4}}\delta_{\lambda -}\end{array}\right);\quad
u_\lambda(p_2)=\left(\begin{array}{c}0\\s^{\frac{1}{4}}\delta_{\lambda
+}\\s^{\frac{1}{4}}\delta_{\lambda -}\\0\end{array}\right),
\label{spinors}
\end{equation}
from which one finds
\begin{equation}
\real\left\{\frac{\langle kp_2\rangle}{[kp_2]}
\frac{[15]^2\langle 25\rangle ^2}{[12]^2}\right\}=
-\frac{(15)(25)}{s}[2\cos^2(\phi-\phi_{2k})-1],
\label{1525}
\end{equation}
with $\phi_{2k}$ the azimuthal relative angle between $k$ and $p_2$.
In the degenerate kinematics of the collinear limit, we may choose to 
absorb $\phi$ in a redefinition of $\phi_{2k}$.
Combining these results with eq.~(\ref{apam}), and reinstating the colour, 
flux, average factors and coupling constants, gives finally
\begin{equation}
\tilde{\cal M}=-\frac{\gs^2\gw^2}{16Ns^2\,u_1^2m_W^2}
(2\cos^2\phi_{2k}-1)(m_W^2-m_t^2)(2m_W^2+m_t^2)(m_W^2m_t^2-ut).
\label{mtilderesult}
\end{equation}
Note that this does indeed give zero contribution to the cross-section when
integrated over the full domain of the azimuthal angle $\phi_{2k}$ relevant
to the branching, as discussed above.


\providecommand{\href}[2]{#2}\begingroup\raggedright\endgroup

\end{document}